\documentclass[review,number,sort&compress,1p,a4paper]{elsarticle}
\usepackage{natbib}
\usepackage{verbatim}
\usepackage[monochrome]{color}
\usepackage{amssymb}
\usepackage{graphicx}
\usepackage{subfigure} 
\usepackage[T1]{fontenc}
\usepackage{ae,aecompl}
\newcommand{\munit}[1]{\ensuremath{\, \textrm{#1}}}

\begin{document}
\begin{frontmatter}

  \title{Design and performance of the South Pole Acoustic Test Setup}
  
  \author[Gent]{Y.~Abdou}
  \author[Wuppertal]{K-H.~Becker}
  \author[Zeuthen]{J.~Berdermann}
  \author[Aachen]{M.~Bissok}
  \author[StockholmOKC]{C.~Bohm}
  \author[Zeuthen]{S.~B\"oser\footnote{Now affiliated with the University of Bonn.}}
  \author[Zeuthen]{M.~Bothe}
  \author[Gent]{M.~Carson}
  \author[Gent]{F.~Descamps\corref{cor}}
  \ead{Freija.Descamps@icecube.wisc.edu}
  \author[Zeuthen]{J-H.~Fischer-Wolfarth}
  \author[Uppsala]{L.~Gustafsson}
  \author[Uppsala]{A.~Hallgren}
  \author[Aachen]{D.~Heinen}
  \author[Wuppertal]{K.~Helbing}
  \author[Zeuthen]{R.~Heller}
  \author[StockholmOKC]{S.~Hundertmark}
  \author[Wuppertal]{T.~Karg}
  \author[Zeuthen]{K.~Krieger}
  \author[Aachen]{K.~Laihem}
  \author[Aachen]{T.~Meures}
  \author[Zeuthen]{R.~Nahnhauer}
  \author[Wuppertal]{U. Naumann}
  \author[Lausanne]{F.~Oberson}
  \author[Aachen]{L.~Paul}
  \author[Zeuthen]{M.~Pohl}
  \author[Berkeley]{B.~Price}
  \author[Lausanne]{M.~Ribordy}
  \author[Gent]{D.~Ryckbosch}
  \author[Aachen]{M.~Schunck}
  \author[Wuppertal]{B.~Semburg}
  \author[Zeuthen]{J.~Stegmaier}
  \author[Zeuthen]{K-H.~Sulanke}
  \author[Zeuthen]{D.~Tosi}
  \author[Berkeley]{J.~Vandenbroucke}
  \author[Aachen]{C.~Wiebusch}

  \cortext[cor]{Corresponding author}

  \address[Gent]{Dept.~of Subatomic and Radiation Physics, University of
    Gent, B-9000 Gent, Belgium}
  \address[Wuppertal]{Dept.~of Physics, University of Wuppertal, D-42119
    Wuppertal, Germany}
  \address[Zeuthen]{DESY, D-15735 Zeuthen, Germany}
  \address[Aachen]{III. Physikalisches Institut, RWTH Aachen University,
    D-52056 Aachen, Germany}
  \address[StockholmOKC]{Oskar Klein Centre and Dept.~of Physics,
    Stockholm University, SE-10691 Stockholm, Sweden}
  \address[Uppsala]{Dept.~of Physics and Astronomy, Uppsala University,
    Box 516, S-75120 Uppsala, Sweden}
  \address[Lausanne]{Laboratory for High Energy Physics, \'Ecole
    Polytechnique F\'ed\'erale, CH-1015 Lausanne, Switzerland}
  \address[Berkeley]{Dept.~of Physics, University of California,
    Berkeley, CA 94720, USA}

  \begin{abstract}
    The South Pole Acoustic Test Setup (SPATS) was built to evaluate
    the acoustic characteristics of the South Pole ice in the 10\,kHz
    to 100\,kHz frequency range, for the purpose of assessing the
    feasibility of an acoustic neutrino detection array at the South
    Pole. The SPATS hardware consists of four vertical strings
    deployed in the upper 500\,m of the South Pole ice cap. The
    strings form a trapezoidal array with a maximum baseline of
    543\,m. Each string has 7 stages equipped with one transmitter and
    one sensor module. Sound is detected or generated by piezoelectric ceramic
    elements inside the modules. Analogue
    signals are sent to the surface on electric cables where they are
    digitized by a PC-based data acquisition system. The data from all
    strings are collected on a central computer in the IceCube Laboratory from where they are send
    to a central data storage facility via a satellite
    link or stored locally on tape. A technical overview of SPATS and
    its performance is presented.
  \end{abstract}
  
  
\end{frontmatter}
\section{Motivation and system requirements}
The sensitive volume needed for the detection of the predicted small
cosmogenic neutrino flux and the study of its angular distribution is
orders of magnitude larger than the instrumented volumes of the
current \v{C}erenkov neutrino telescopes~\cite{Achterberg:2006md,Aguilar:2006rm}. New detection
methods that are sensitive to the radio or acoustic signatures of a
UHE neutrino interaction would allow a more sparse instrumentation and
therefore a larger sensitive volume at reasonable
cost~\cite{Kravchenko:2002mm,Thompson:2008zz,Gorham:2010kv,Aguilar2011128}. The feasibility and
specific design of an acoustic array as part of a hybrid
optical/radio/acoustic neutrino detector, first suggested and
simulated in \cite{Besson:2005re}, depends upon the acoustic
properties of the South Pole ice in the concerned frequency region
(10--100)\,kHz. The South Pole Acoustic Test Setup (SPATS) was built
to evaluate the attenuation length, speed of sound, background noise
level and transient noise rate of the South Pole ice and was deployed
in the 2006/2007 and 2007/2008 polar seasons.

\section{Geometry}
\begin{figure}
 \centering
 \includegraphics[width=\textwidth]{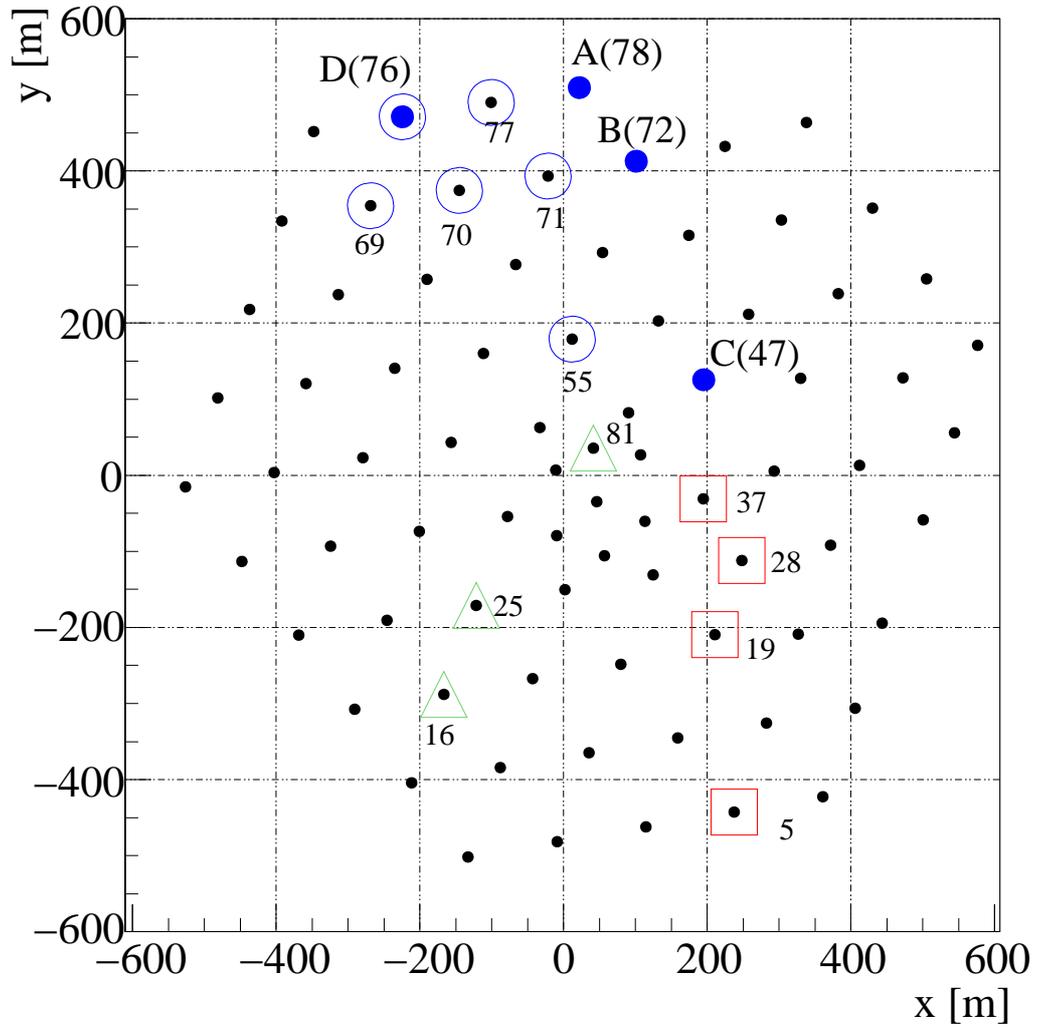}
 \caption[The SPATS geometry.]{SPATS strings (large dots) overlayed
   on the IceCube geometry as of February 2009 (small dots). The
   String\,ID (A,B,C or D) is given followed by its corresponding
   IceCube hole number. The open circles, open squares and
   open triangles show the positions of the
   2007/2008, 2008/2009 and 2009/2010 pinger holes (the \emph{pinger}
   is a retrievable acoustic transmitter, see~\S\,\ref{sec:pinger}) with the
   corresponding IceCube hole number.}
\label{fig:SpatsGeometry}
\end{figure}

\begin{figure}[h]
 \centering
 \includegraphics[width=\textwidth]{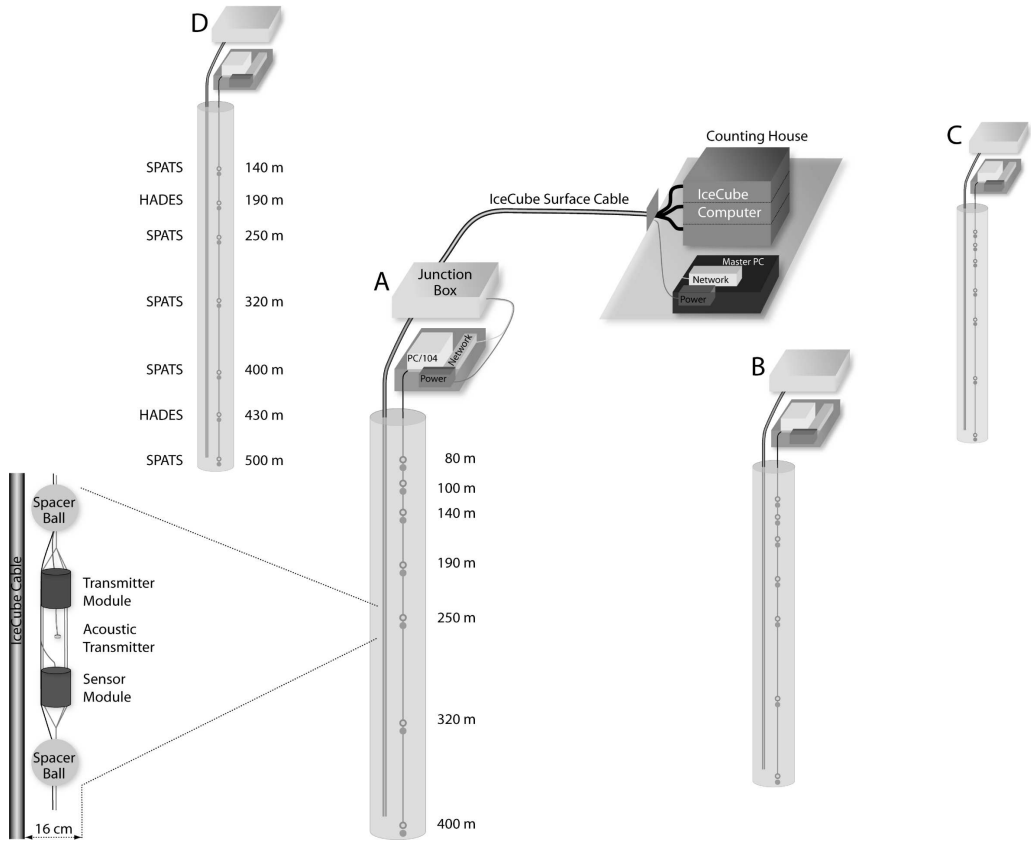}
 \caption[Schematic of the SPATS array.]{Schematic of the SPATS array,
   with the four strings consisting of seven acoustic stages.}
\label{fig:SpatsLayout}
\end{figure}

The permanently installed hardware of SPATS consists of four
instrumented cables, called strings, that were deployed
vertically in the upper 500\,m of selected
IceCube~\cite{Achterberg:2006md} boreholes to form a trapezoidal array.
The current geometrical configuration (see
Fig.~\ref{fig:SpatsGeometry} and Fig.~\ref{fig:SpatsLayout}) is the
result of a compromise between the geometry necessary to achieve the
physics goals and the availability of IceCube holes and personnel
availability during deployment.

Measuring, for example, the acoustic properties of the ice both
parallel and perpendicular to the flow of the glacier permits the
investigation of possible anisotropies of the acoustic properties of
the ice. Therefore good horizontal coverage is needed. For an
attenuation length analysis, it is important to have sufficient
variation in transmitter to sensor distances so that the fit of
amplitude versus distance is well constrained. The uncertainty on the
horizontal position of each string is fixed and known to be
$\pm0.5$\,m, so that the relative error decreases with increasing
string-to-string distance (so called baseline). The
vertical distance between the acoustic transducers was chosen to
increase with depth following the measured and predicted temperature
and density profiles of the
ice~\cite{Price:2002,Albert}. Table~\ref{tab:deploy} gives the
deployment details and baselines for SPATS.

\begin{table}
  \centering
  \begin{tabular}{c|c|c|c}
    \hline\hline
    String & Deployed  & Baseline~(m) & Instrumented depths~(m)\\
    \hline\hline
    A & 14$^{th}$~January~07 & (A-B) 125 & 80,100,140,190,250,320,400\\
    B & 11$^{th}$~January~07 & (B-C) 302 & 80,100,140,190,250,320,400\\
    C & 22$^{nd}$~December~06 & (C-A) 421 & 80,100,140,190,250,320,400\\
    D & 24$^{th}$~December~07 & (D-C) 543 & 140,190(H),250,320,400,430(H),500\\
    \hline
  \end{tabular}
  \parbox{12cm}{\caption{Deployment details for SPATS, the two HADES levels on String\,D are indicated.}\label{tab:deploy}}
\end{table}

Figure~\ref{fig:SpatsLayout} shows a schematic of the SPATS array with
its in-ice and on-ice components. Each of the four strings has seven
acoustic stages at specified depths. Table~\ref{tab:deploy} shows the
corresponding levels for
each string. It was decided to leave out the two most shallow
instrumented levels on String\,D after the 2007 data-set hinted towards
less optimal acoustic conditions in those upper levels (higher
background noise and shorter attenuation length). String\,D has
second-generation SPATS transmitters and sensors. In addition, the
Hydrophone for Acoustic Detection at South Pole (HADES,
see~\cite{Semburg:2008ce}) was deployed on two levels of String\,D (190\,m and 430\,m). The
acoustic modules are discussed in more detail in the following
sections.

\section{In-ice components}
\begin{figure}
 \centering
 \includegraphics[]{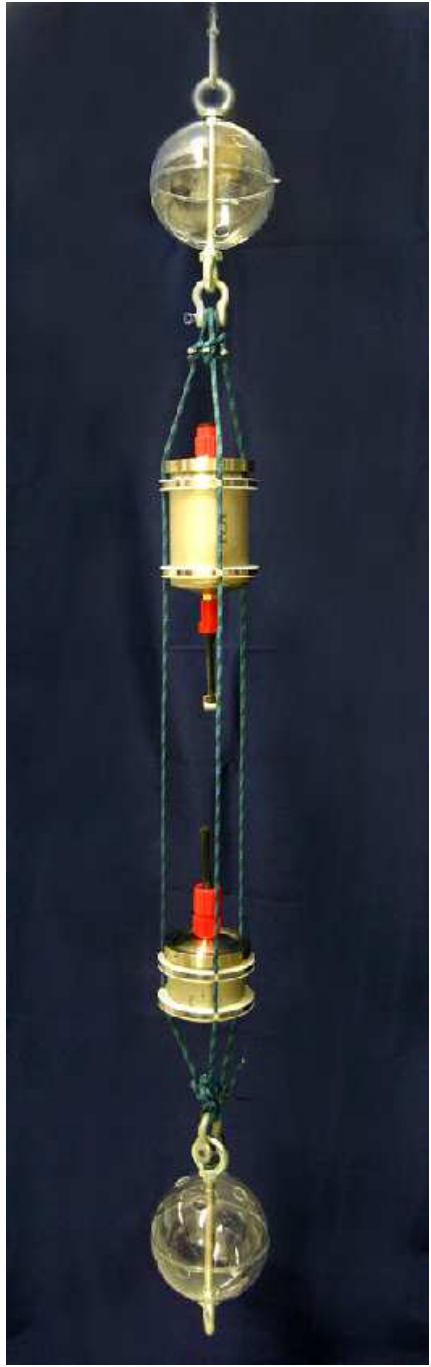}
 \caption[]{Fully assembled SPATS stage.}
\label{fig:stage}
\end{figure}

Each acoustic stage consists of a separate transmitter and sensor
module. All the electronic circuits are located in
steel\footnote{Stainless steel grade 304/1.4301} pressure housings
which have an outer diameter of 101.6\,mm and a wall thickness of
5.74\,mm. The transmitter and
sensor modules are joined together with three ropes, allowing the
transmitter module to be mounted about 45\,cm above the sensor
module (see Fig.~\ref{fig:stage}). The ropes are attached at the top and bottom to a steel nut
that is threaded and mounted on a bolt. This bolt also runs through a plastic
perforated hollow sphere. The holes allow for water to flow in so that
the spheres remain intact at least until freeze-in. These spacer balls
are 16\,cm in diameter and assure a minimum distance of the stage to
the IceCube main cable and the wall of the hole. An entire stage is
about 1.5\,m long and maximum 16\,cm wide with a total weight of
10\,kg.

Each acoustic stage is connected to an Acoustic Junction Box (AJB),
located at the surface, by two shielded cables (one for each module)
each consisting of four twisted pairs. All in-ice cables are threaded
around a support rope which carries the weight of the stages. The
depth of the string was monitored during deployment by
reading out the pressure sensor that is inside each of the bottom-most
transmitter modules. The resulting precision in the depth measurement is 
estimated to be $\pm$2\,m.

\subsection{The transmitter module}
The SPATS transducers acoustic elements all consist of the same lead zirconium titanate
(PZT) material, namely PIC151, manufactured by PI-Ceramics. This is a
soft piezo-ceramic material with a high piezoelectric charge constant
(d$_{33}$=500$\frac{pC}{N}$), high permittivity and a high coupling
factor.

The transmitter module consists of a steel pressure vessel that houses
a high-voltage pulse generator board (hereafter HV-board) and a temperature sensor. For the
deepest stages, the temperature sensor is replaced with a commercial
pressure sensor. Both sensors give a linear output current range of
(4--20)\,mA, which translates to a temperature range of
--70\,$^{\circ}$C to +10\,$^{\circ}$C and a pressure range of
(0--60)\,bar. The HV-board hosts a a circuit
designed to pulse the piezoelectric ceramic with a high voltage
pulse. An LC circuit is used as an intermediate stage to locally store
the charge necessary for the generation of the HV pulse. The total
charge accumulated is determined by two remotely controllable
parameters: the length of the TTL trigger signal during which the LC circuit is charged
and the DC steering
voltage. When the TTL pulse goes to a high state, a current (defined
by the DC steering voltage), flows across the LC stage. When the TTL
pulse goes to a low state, the charge accumulated is discharged onto
the piezo and an acoustic pulse is emitted. The charging voltage is
variable and is provided by a voltage regulation board. 

\begin{figure}
  \centering
  \subfigure[]{\label{fig:HVRB1}\includegraphics[height=5.5cm]{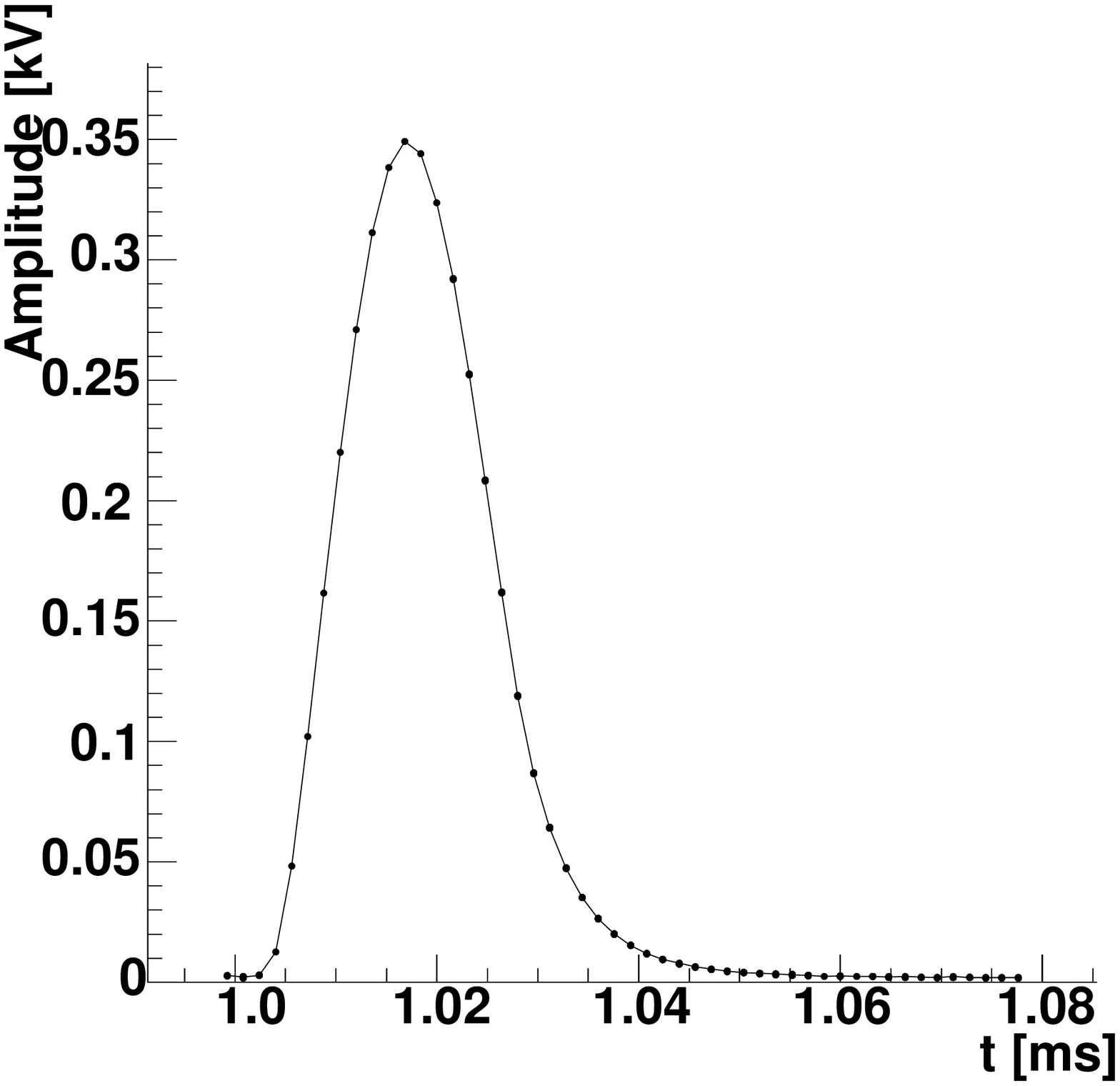}}
  \subfigure[]{\label{fig:TBFourier}\includegraphics[height=5.5cm]{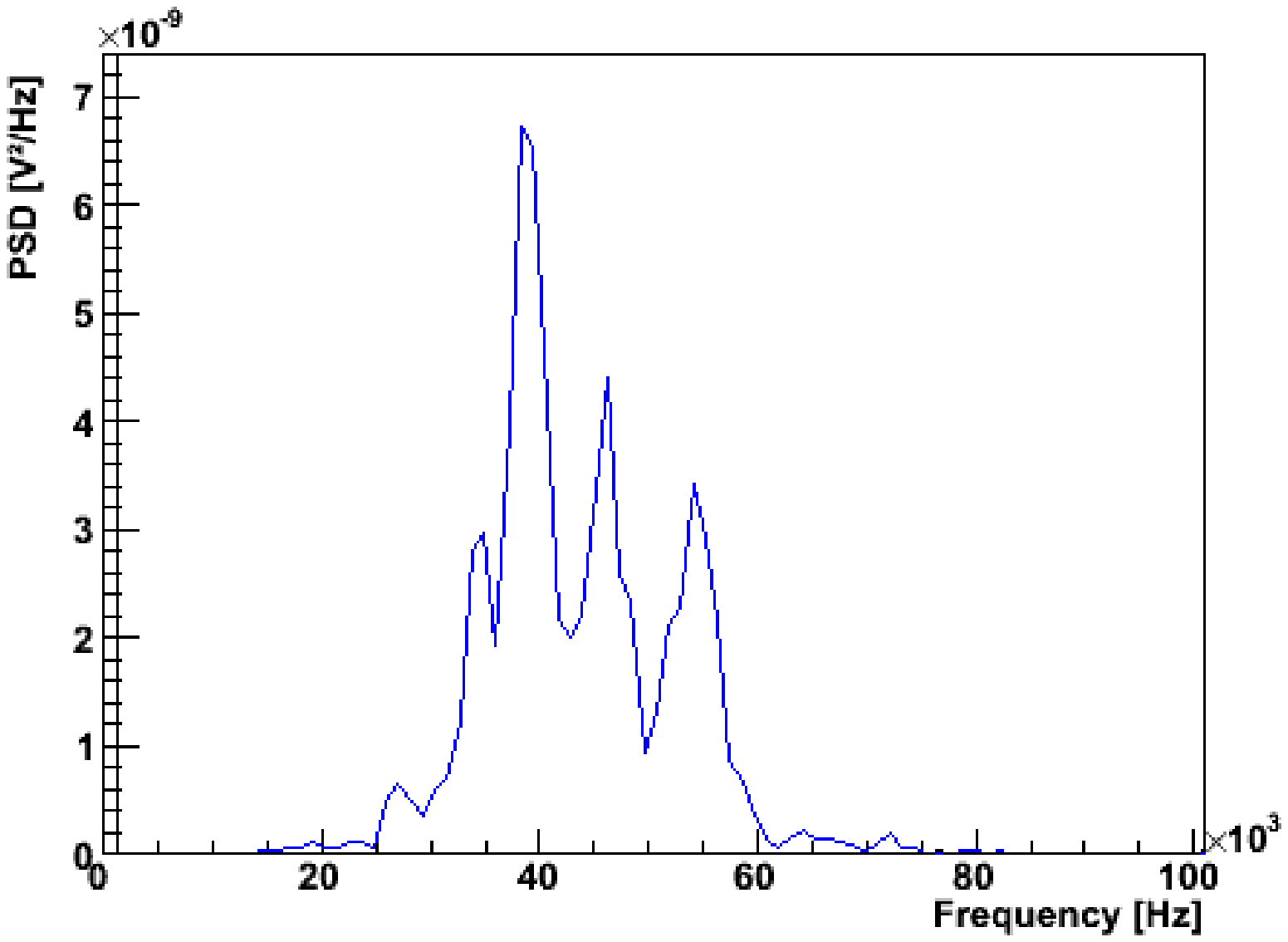}}
  \caption[Transmitter module high voltage pulses.]{(a): An averaged high voltage pulse for the first generation SPATS transmitter. The statistical error bars are too small to be seen on this figure. (b): Fourier spectrum
    for a String\,B transmitter as seen by the HADES sensor at 430\,m
    depth.}
\end{figure}
Gaussian shaped HV pulses with a FWHM of about 17\,$\mu$s
and a variable maximum up to 1.5\,kV are sent to the piezo-ceramic
element. The resulting
HV pulse is read back as a differential signal (HVRB). This allows
effects of cable length and temperature to be studied and taken into
account in subsequent analyses. Figure~\ref{fig:HVRB1} shows an 
averaged HV pulse, averaged over ten individual pulses, generated with a first-generation SPATS transmitter
module. 
This shows a very high
repetition stability. These pulses are triggered by TTL signals of
5\,ms duration. Figure~\ref{fig:TBFourier} shows the Fourier spectrum
for a waveform from HADES (see~\S\,\ref{sec:hades}) listening to a
String\,B transmitter (BT4). The SPATS transmitters typically output most
power in the 20\,kHz to 60\,kHz range, although the variations of the
actual shape of the emission spectra are large.
The String\,D transmitters have an optimized HV-circuit design which
results in sinusoidal half-wave HV pulses with a FWHM of
about 50\,$\mu$s and higher pulse amplitudes. These pulses are
triggered by TTL signals of typically 2\,ms duration.

The ring-shaped piezo-ceramic element is cast in epoxy for electrical insulation and
positioned $\sim$13\,cm below the steel housing. Azimuthally isotropic
emission is the motivation for the use of ring shaped
piezo-ceramics. The actual emission directivity of such an element was
measured in azimuthal and polar directions,
see~\S\,\ref{sec:transdircal}. The ring-shaped ceramic element is not
isotropic in the polar plane. The HV cable that connects this element
to the transmitter module has limited flexibility and an extra
rigidity was added before deployment to avoid bending. Still, it can
be expected that the piezo has a certain angle to the horizontal
plane. In this case the amplitude variation in the horizontal plane
will be larger than expected from a perfectly horizontally mounted piezo.

\subsection{The sensor module}
The sensor module has three channels placed 120$^{\circ}$ apart in
order to ensure good angular coverage. A channel consists of a
cylindrical (10\,mm diameter and 5\,mm height) piezo-ceramic element
of the PZT type that is pressed against the steel housing. The ceramic
element is directly soldered to a 3-stage amplifier, the first stage
of which consists of a low-noise amplifier\footnote{AD754JR} with an
amplification factor of 100, a large bandwidth and a 3\,kHz high-pass
filter to suppress low frequencies. The second amplifier stage is
an AC coupled inverted amplifier that amplifies the signal by a factor
of 100 and suppresses frequencies above 100\,kHz. The last stage is a
line driver: it prepares the signal for transport over the long in-ice
cable by providing a differential output with a gain of 1. Two twisted
cables then transport the complementary signals to the differential
input of the ADC in the string-PC. This reduces noise by
rejecting common-mode interference. The total amplification factor of
the chain of amplifiers is of the order of 10$^{4}$. 

The 21 first-generation sensor modules that were deployed on strings
A, B and C have a central bolt that is connected to three
screws. These preload screws put pressure on the ceramic element
through the amplifier board to ensure good contact with the steel
housing and avoid deformation of the module. This introduces a
mechanical coupling between the three different channels of the sensor
module. It was therefore decided to replace the preload screws with a
metal ring in the second-generation sensor modules on String\,D.
\begin{figure}[h]
  \centering \subfigure[SPATS first generation sensor
    module.]{\label{fig:sensormodule1}\includegraphics[height=4.0cm]{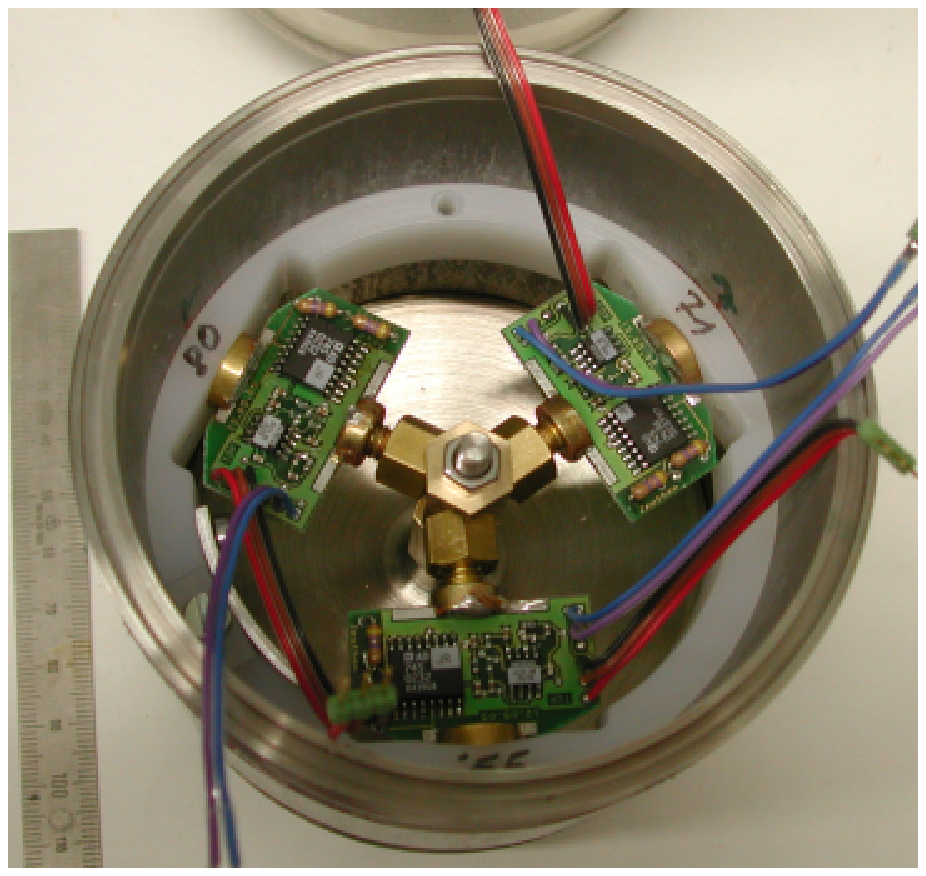}}
  \hspace{0.2cm}\subfigure[SPATS second generation sensor
    module.]{\label{fig:sensormodule2}\includegraphics[height=4.0cm]{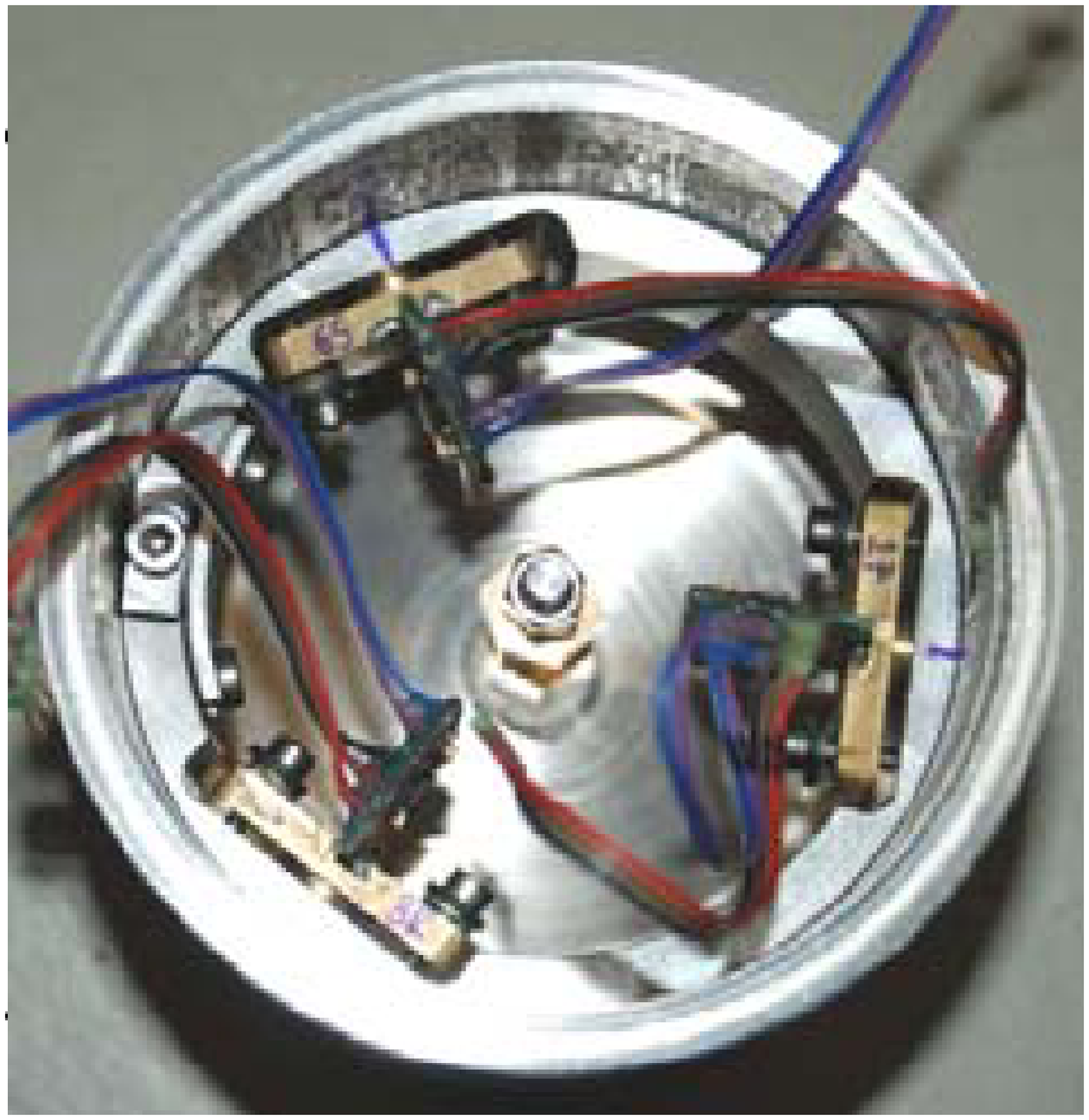}}
  \hspace{0.2cm}\subfigure[HADES
    sensor.]{\label{fig:hadesmodule}\includegraphics[height=4.0cm]{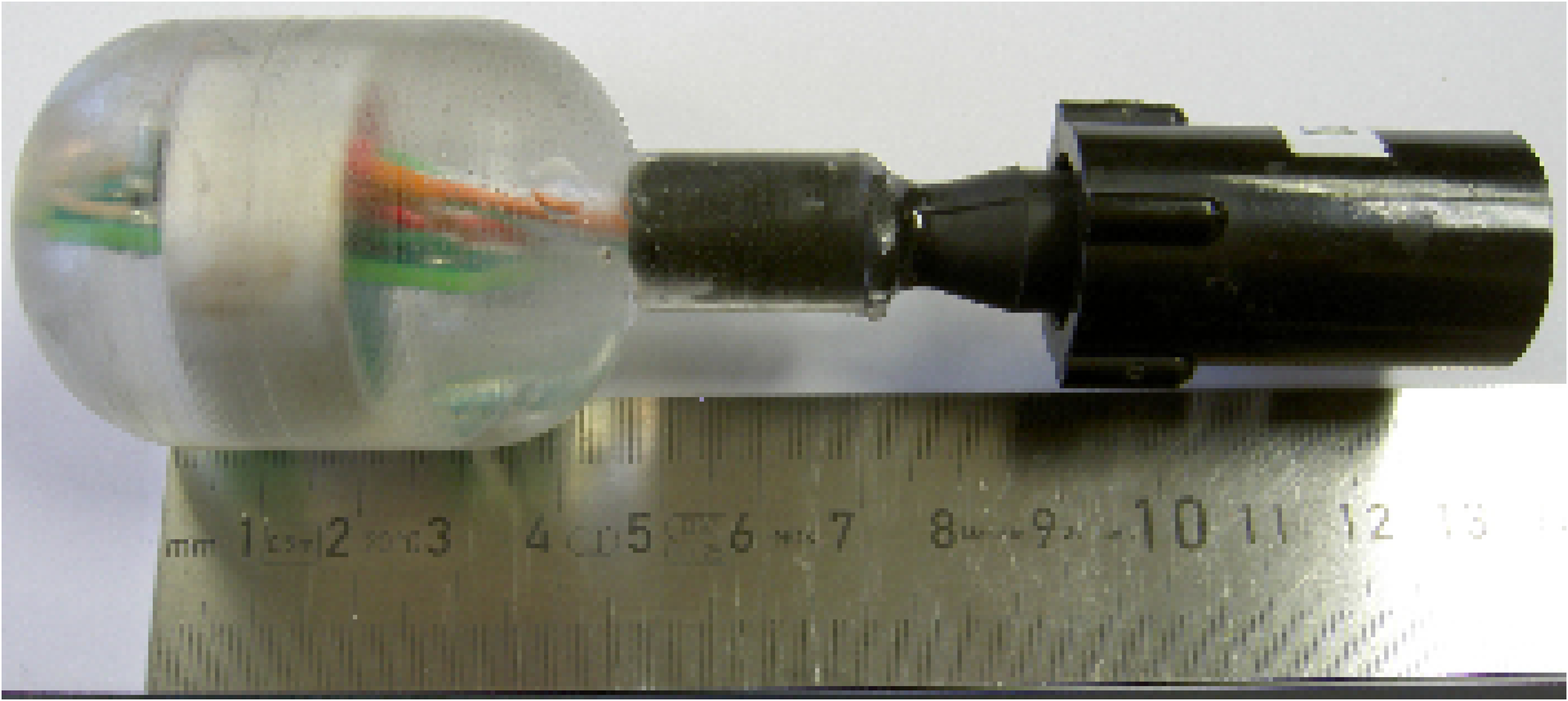}}
  \caption[The SPATS and HADES sensors.]{Pictures of the SPATS and
    HADES sensors. The SPATS first (a) and second (b) generation
    sensor modules have 3 channels consisting of a piezo-ceramic
    element followed by an amplifier board. The HADES sensor (c)
    consists of a cylindrical piezo-ceramic element with a amplifier
    board cast in resin.}
  \label{fig:sensormodules}
\end{figure}
Figure~\ref{fig:sensormodules} shows pictures of open first
(\ref{fig:sensormodule1}) and second (\ref{fig:sensormodule2})
generation SPATS sensor modules.

\subsection{HADES}\label{sec:hades}
The Hydrophone for Acoustic Detection at South Pole (HADES) was
developed in order to offer an alternative in-ice sensor with a
different dynamical range. HADES-A and HADES-B were 
deployed on String\,D at depths of 190\,m and 430\,m. A ring-shaped
piezo-ceramic element is connected to a two-stage differential
amplifier that is placed inside the ring. The assembly is then coated
with resin (two component hard polyurethane). The choice of resin was
made after a series of laboratory tests that investigated the
resistance of the material to temperatures down to
-85~$^{\circ}$C. Also, the acoustic impedance matching was calculated
from the measured sound speed for each
material. Figure~\ref{fig:hadesmodule} shows a picture of the HADES
sensor.

\subsection{Naming conventions}
The SPATS strings were named String\,A, B, C and D. The three sensor
channels of each sensor module are each read out by a different fADC
board. Therefore the sensor channels are named channel 0, 1 or 2,
based on the fADC board to which they are connected. A sensor channel is
named {\em X}\,S{\em Y}-{\em Z}, pointing to the sensor channel {\em Z}
of the sensor module at position {\em Y} on String\,{\em X}. The naming
convention for a transmitter is {\em X}\,T{\em Y}: the transmitter on
String\,{\em X} at position {\em Y}. Stage positions {\em Y} are number
1 to 7 from the top of the string to the bottom. Channel 0 of the top sensor of 
String\,A is designated as AS1-0. 

\subsection{Pinger}\label{sec:pinger}
In addition to the permanently installed instrumentation which has
been described above, a retrievable transmitter, called {\em pinger},
was used at the South Pole in the 2007/2008, 2008/2009 and 2009/2010
seasons. A description of the pinger can be found in
\cite{Collaboration:2009sia} and \cite{Abbasi:2010vt}.

\section{On-ice components}
\subsection{The acoustic junction box}
\begin{figure}
 \centering \subfigure[A view of the
   AJB]{\label{fig:ajb}\includegraphics[height=5cm]{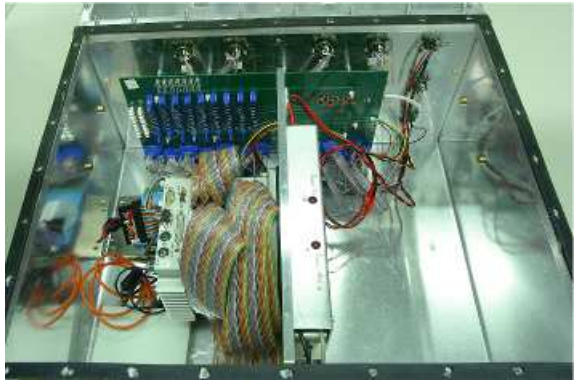}}
 \hspace{0.5cm}\subfigure[The string-PC with DSL and serial
   modem.]{\label{fig:string-pc}\includegraphics[height=5cm,
     width=4 cm]{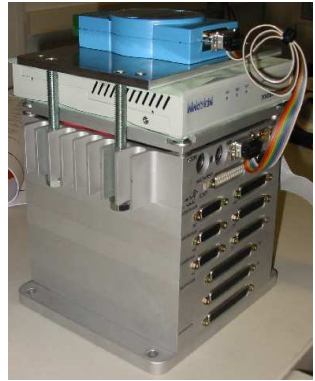}}
  \caption[The acoustic junction box (AJB) and string-PC.]{(a): The
    watertight part of the acoustic junction box (AJB) with the PCB,
    power distribution board, string-PC and modems. (b): The
    string-PC stack. From top to bottom are the serial and DSL modems,
    the CPU board, 3 fast-ADC boards, 1 slow-ADC board and 1 relay
    board.}
\end{figure}

The AJB is a robust aluminium box (dimension:
$(30\times50\times80)$\,cm for strings A, B and C and slightly larger
for String\,D) that is situated above each acoustic string and buried
under roughly 3\,m of snow. It is split into two compartments; one
holds the connectors from the in-ice cables. These are connected to wall-mounted
connector-sockets (patch wall). The other compartment is sealed
watertight and contains the electronic components (see
Fig.~\ref{fig:ajb}). All in-ice signals
are first routed through a printed circuit board (PCB) from where they
are distributed to the DAQ-boards of a low-power industrial PC, called
the string-PC. A power distribution unit (PDU) consists of low-noise
DC/DC converters and filters. It provides ground, +5\,V, +15\,V and
+24\,V to the string-PC and the in-ice components through the PCB. The
PCB also routes the GPS timing signal and converts the currents from
the in-ice pressure and temperature sensors to voltages.
\subsection{The string-PC}
The string-PC is a stack of IDAN (Intelligent Data Acquisition Node)
PC/104 modules by RTD\footnote{http://www.rtd.com/} (see
Fig.~\ref{fig:string-pc}). It has a compact
modular design and its splash-proof rugged aluminium enclosure acts as
a heat sink so that no fan is required. It is rated to perform from
--40\,$^{\circ}$C to +85\,$^{\circ}$C. The peripheral components are
controlled by a CPU module\footnote{IDAN-CML47786HX650ER-260D} with a
600\,MHz processor and 512\,MB RAM. The Linux operating system is
installed on a flash-memory (wide temperature disk-on-module
(DOM)\footnote{DJ0010G44TK02P10; operating temperatures:
  --40\,$^{\circ}$C to +85\,$^{\circ}$C
  (http://www.pqimemory.com/)}). This type of solid-state memory has
the advantage that there are no moving parts, unlike a regular
hard drive. The drawback is that each memory-cell of the DOM can be
accessed for writing a limited number of times. If too many
cells fail the disk will become unaccessible. For this reason, all
runtime data are stored locally on a temporary RAM-disk. This means
that part of the memory of the CPU module is treated as if it were a
disk drive. Hence, the read-write cycles to the DOM are kept to a
minimum.  The 12-bit, 8 channel fADC boards\footnote{IDAN-SDM7540HR-8}
have a maximum sampling frequency of 1.25\,MHz. There are three fADCs
for each string, which means that all three channels of one sensor
module can be read out simultaneously at a maximum sampling frequency
of 1.25\,MHz. Theoretically, the fADCs should be able to read out all
21 sensor-channels of the string at maximum $\sim$179\,kHz sampling
frequency. The three boards are linked through a SyncBus (RTD)
connection which allows the simultaneous recording of samples (using
the same ADC clock of one of the boards). Together with one slow
ADC/DAC board\footnote{IDAN-DM6420HR-1-62S} with a sampling frequency
of 500\,kHz, the transmitters can be controlled and the temperature
and pressure sensors can be read out.

A relay board\footnote{IDAN-DM6952HR-62D} has 16 relays which allows
the power for each sensor, transmitter and the power to the temperature and 
pressure sensors to be switched on and off individually.
The average power
consumption per string is low and varies from $\sim$35\,W (no in-ice
modules powered on) to $\sim$96\,W (all in-ice modules
powered-on). All these components were tested at the expected low
temperatures (around --55\,$^{\circ}$C) and several cold boot cycles
of the system were successfully performed prior to deployment. The
string-PCs have been powered-on without problems after extended power-outages
at the South Pole.

The AJB allows for over 90 analogue channels to be controlled and
read-out without the need to transport the analogue signals over
dedicated surface cables to the IceCube Laboratory (ICL). Signal
losses and cost are therefore kept to a minimum. Each AJB is connected
to the master-PC located in the ICL by two cables (quads) of the
IceCube surface cable. They each have 2 twisted wire-pairs: DC-power
and SDSL\footnote{A symmetric DSL connection has the same bandwidth for
  the upstream and downstream connections.} communication on one quad
and DC-power and a GPS signal on the other. All patch cables are
shielded and the shield is connected to the AJB shield. The DC
power is automatically inhibited in hardware and firmware if the cable
shields are not intact.
\begin{figure}[t]
 \centering
 \includegraphics[width=\textwidth]{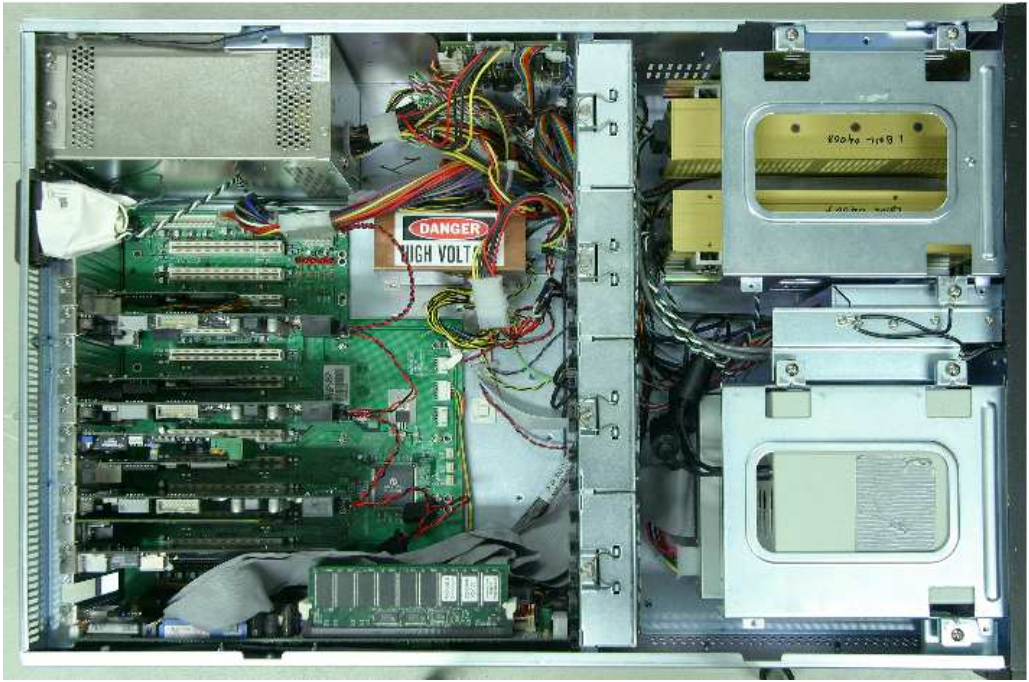}
 \caption[Top view of the master-PC.]{Top view of the master-PC.}
\label{Chap3_masterpc}
\end{figure}

\subsection{The master-PC}
The master-PC (Fig.~\ref{Chap3_masterpc}) is a rackable 4U HP Proliant
DL380 series with dual 1GHz Pentium-3 processors and a 12-slot PCI
backplane\footnote{Retrofitted IceCube Rev 1 DOM Hub.}.  Two
Acopian\footnote{W48NT370, http://www.acopian.com/} switching
regulated power supplies connected in series supply
96\,V$_{\textrm{DC}}$. They are cooled independently by internal
dual fans.  A symmetric DSL connection is assured by Nexcomm Nexpeed
NM220G modems. These SHDSL modems provide symmetrical data rates of a
maximum of 2.3\,Mbps for a distance of up to 3\,km over a regular
wire-pair. In-situ testing has shown the baseline ping-delay to be
around 1.5\,ms for all four strings. A GPS-based IRIG-B (100\,pps)
time coding signal, provided by a Meinberg GPS PCI card (GPS169PCI)
guarantees absolute time stamping. The GPS clock is specified to
produce IRIG-B rising edges within $\pm$2\,$\mu$s of absolute GPS
time. The delay introduced in the IRIG-B signals during propagation
from master-PC to string-PC is a few $\mu$s.  Each string has a
dedicated PCI control board (SPATS Hub Service Board: SHSB) and
ethernet card. The SHSB routes the DSL communication, GPS signal and
power to the two surface cable quads. The SHSB firmware can be used to
power on and off the strings, as well as to monitor the status of the
wire pairs. All minimum and maximum voltages and currents can be set
individually and the power is automatically cut-off the instant one of
the thresholds is crossed. 
The master-PC is accessible through the local ICL network. Remote
access to that network is only possible when the satellite link with
the South Pole Amundson-Scott station is active. Currently, all
communication and data-transfer for the entire station are executed
over satellites,\footnote{For satellite times, see for example
  http://ice.rsmas.miami.edu/access.phtml} with a total of eight hours
of visibility. All SPATS-data is immediately transferred from the
string-PCs to the master-PC where it is compressed and stored until
transferred. A specialized shell script then prepares different data
streams for both satellite transfer and tape archiving. Data transfer capabilities are very
limited at the South Pole. SPATS has
been assigned a maximum of 150\,MB of satellite-transferred data per
day. It can take up to five days for the data to arrive on the IceCube
data-servers after the data is picked-up by the satellite system. All
data that is not transferred automatically goes to tapes that are
brought back from South Pole every year.

In addition to the SPATS system at the South Pole, two Northern
Hemisphere test-systems also exist: String\,Z and String\,E. String\,Z
consists of a complete string-PC that is connected to a desktop-PC
(simulating the master-PC) through an SDSL connection. It is possible
to connect SPATS sensors, transmitters or a function
generator. String\,E consists of a CPU-board and an fADC board to
which a function generator can be attached directly. Both test-systems
have been extensively used for DAQ software development.

\section{System performance}
\label{sec:fadc}
\subsection{fADC performance}
\begin{figure}[t]
  \centering
  \includegraphics[width=\textwidth]{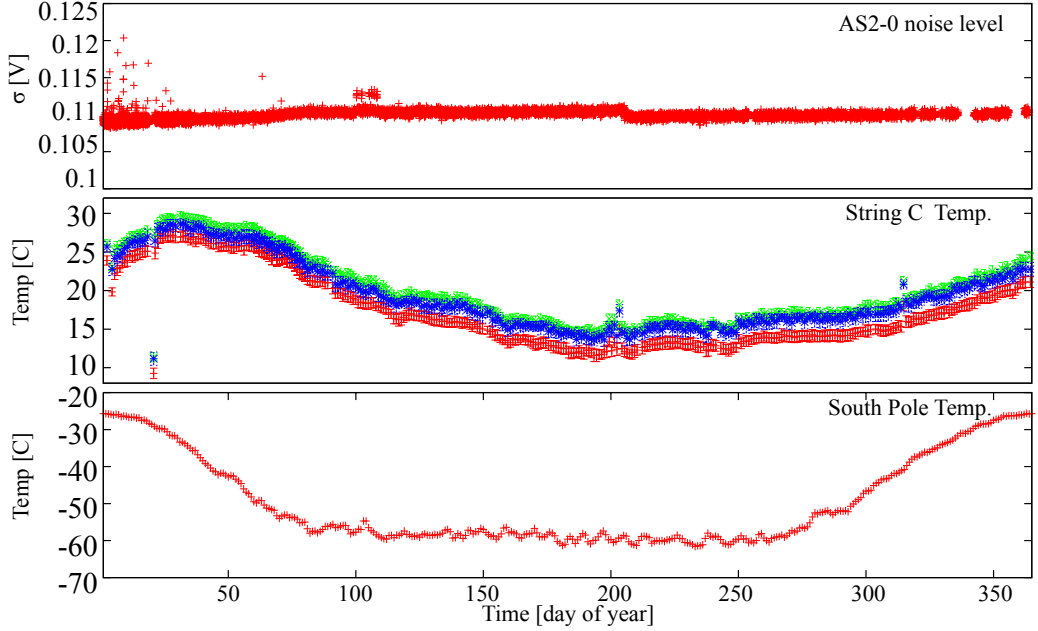}
  \caption[Evolution of the ADC temperatures and noise levels compared
    to the average surface temperature at the South Pole.]{Evolution of
    the fADC temperatures and AS2-0 noise levels compared to the average
    surface temperature at the South Pole. The AS2-0 sensor channel is
    located at a depth of 100\,m, at the interface between the firn and the bulk ice.}
  \label{fig:adctemps}
\end{figure}

The temperature of each fADC board is influenced by both external
(weather, station-wide power outages) and internal (data-taking modes)
factors. The fADC temperatures (and therefore the AJB temperature)
take about 12 hours to stabilise after an extended power outage. Over
the course of an entire year, the fADC temperature changes by a
maximum of 20\,$^{\circ}$C whereas the South Pole surface temperatures
fluctuates by more than 35\,$^{\circ}$C. The insulating snow that surrounds
the AJB acts as a buffer to the extreme temperatures: the temperature
fluctuations are smaller and have a delay of about 1 month, see
Fig.~\ref{fig:adctemps}.

Each fADC has a maximum sampling frequency of 1.25\,MHz. The fADC
clocks were found to drift at a rate that is typically several parts
per million, this means that the fADC will not sample the waveform at
the requested (nominal) frequency but a slightly higher or lower
one. The cumulative amount of drift causes decoherence by
averaging of SPATS transmitter and pinger pulses if the nominal rather
than true sampling frequency is used. This clock drift effect can be
accounted for by using the IRIG-B GPS signal since the latter is
recorded simultaneously with each sensor channel recording. It is
therefore possible to determine the actual sampling frequency at the
time of the recording. This is the sampling frequency used for
subsequent analyses.

\subsection{Current system status}
The status of the sensor channels can easily be assessed by
investigation of the ADC-count histogram for noise.  Currently 6 out
of a total of 80 sensor channels are continuously saturated (AS3 and
CS1) and disregarded for analysis. The status of the transmitters can be assessed by looking at
their HVRB. Most of the transmitters from strings A, B and C have
evolved after their deployment in the ice from displaying smooth
HVRB-pulses to more structured ones. The pulses remain very
reproducible although the amplitudes typically have
decreased. Currently, all but one of the SPATS transmitters are
operational.

\section{Data Acquisition}
\subsection{Data format}
All strings run exactly the same software at
any given time. An electronic logbook keeps track of all activity on
the master and string-PCs.

All waveform data is recorded in a binary format and is automatically assigned an
incremental run-number. A SPATS run consists of a certain number
of events that can each contain a sensor waveform and/or transmitter
HVRB waveform. A waveform consists of samples that contain both the
IRIG-B and ADC counts so that each sample has precise time stamping.

\subsubsection{System monitoring data-taking}
The monitoring data allows the basic system parameters to be
controlled for both the master-PC and the string-PCs. The string-PC
monitoring script runs once per hour and collects the following data:
\begin{enumerate}
 \item PTS monitoring: the pressure and temperature sensors inside the
   transmitter modules are read,
 \item fADC temperatures: the internal temperature sensor is read-out
   for all fADC boards,
 \item Network Time Protocol (NTP) variables. Jitter on the system
   time NTP synchronisation is monitored. The jitter is a measure of
   the phase noise in the time received from the server, if the time
   queries happen at uneven intervals, the jitter is high,
 \item The string-PC DOM and RAM-disk status.
\end{enumerate}
The daily runlog keeps track of the start- and stop-times of all runs.
The master-PC monitoring script runs each hour and collects the
following data:
\begin{enumerate}
 \item power monitoring: the status of the wire pairs for all SPATS
   surface cables is monitored,
 \item ping-delays: the average ping time between the string-PCs and
   the master-PC is recorded,
 \item the master-PC hard disk status,
 \item the status of the NTP sever on the master-PC.
\end{enumerate}
The results of the system monitoring data-taking are compiled in an email
(monitor-mail) and sent over the Iridium cellular satellite network
during satellite downtimes so that action can be taken right from the
start of the next satellite pass. In addition, all data-transfer
details are compiled in another email. Herein, the total data-transfer
rate for that day and the amount of data that went to the local taping
system, without being transferred to the north, are given.

\subsubsection{Acoustic data-taking}
There are two main C-programs that can be used for a variety of
acoustic data-taking modes. In most modes, the three channels of each
sensor module are recorded simultaneously and the data-taking loops
over all stages of the string. This scheme allows each sensor channel
to be recorded at a sampling frequency of 200\,kHz. The data taking modes are:
\begin{itemize}
 \item {\em Untriggered noise}. The noise waveform is recorded for a short
   duration (currently 0.1\,s) at a sampling frequency of 200\,kHz. These
   full waveforms are used for the absolute noise level analysis, since
   they retain all frequency information up to 100\,kHz.
 \item {\em Triggered noise}. If the number of ADC counts on any of the
   twelve monitored channels (three on each string: one for each ADC
   board) exceeds a certain level above noise, we record a 5\,ms
   window of data (1001 ADC samples at 200\,kHz) centered on the
   trigger in that channel. The resulting trigger rate is roughly
   stable and of the order of a few triggers every minute for each of
   the monitored channels. Most of these events are Gaussian noise
   events, where only one sample is outside the trigger
   boundaries. The transient events are processed off-line and
   analysed for time-coincidence clustering. This also allows
   monitoring of the fADC clock drifts and the recording of
   noise-histograms during transient data-taking.
 \item {\em Untriggered noise histograms}. It was found that it was
   desirable to have a way of monitoring the noise level for all
   sensor channels without having to save the entire waveform for each
   channel. Indeed, this last option uses a large amount of
   disk space and SDSL transfer time. Therefore, the noise is additionally recorded
   in the form of an ADC-count histogram.
 \item {\em Intra-stage, intra-string and inter-string}. A sensor
   records the acoustic pulses originating from the transmitter on the
   same stage, same string or on a different string. In a typical
   inter-string data-taking schedule, one transmitter is triggered
   while all of the other strings loop over their sensors. This way, all
   transmitter-sensor combinations are obtained.
 \item {\em Pinger runs}. A sensor records the acoustic pulse
   originating from the retrievable transmitter. For the 2008/2009 pinger season, the
   pinger signal is recorded for 18\,s at a sampling frequency of
   200\,kHz by all three channels of the sensor module
   simultaneously. A string completes a loop over all sensor modules in
   less than 4\,min. The 4 SPATS strings can record the
   same module at the same time within 10\,ms due to the NTP
   synchronisation and the fact that the data-taking script is
   restarted every 4\,min.

\end{itemize}

\begin{figure}
\begin{center}
\includegraphics[width=0.9\textwidth]{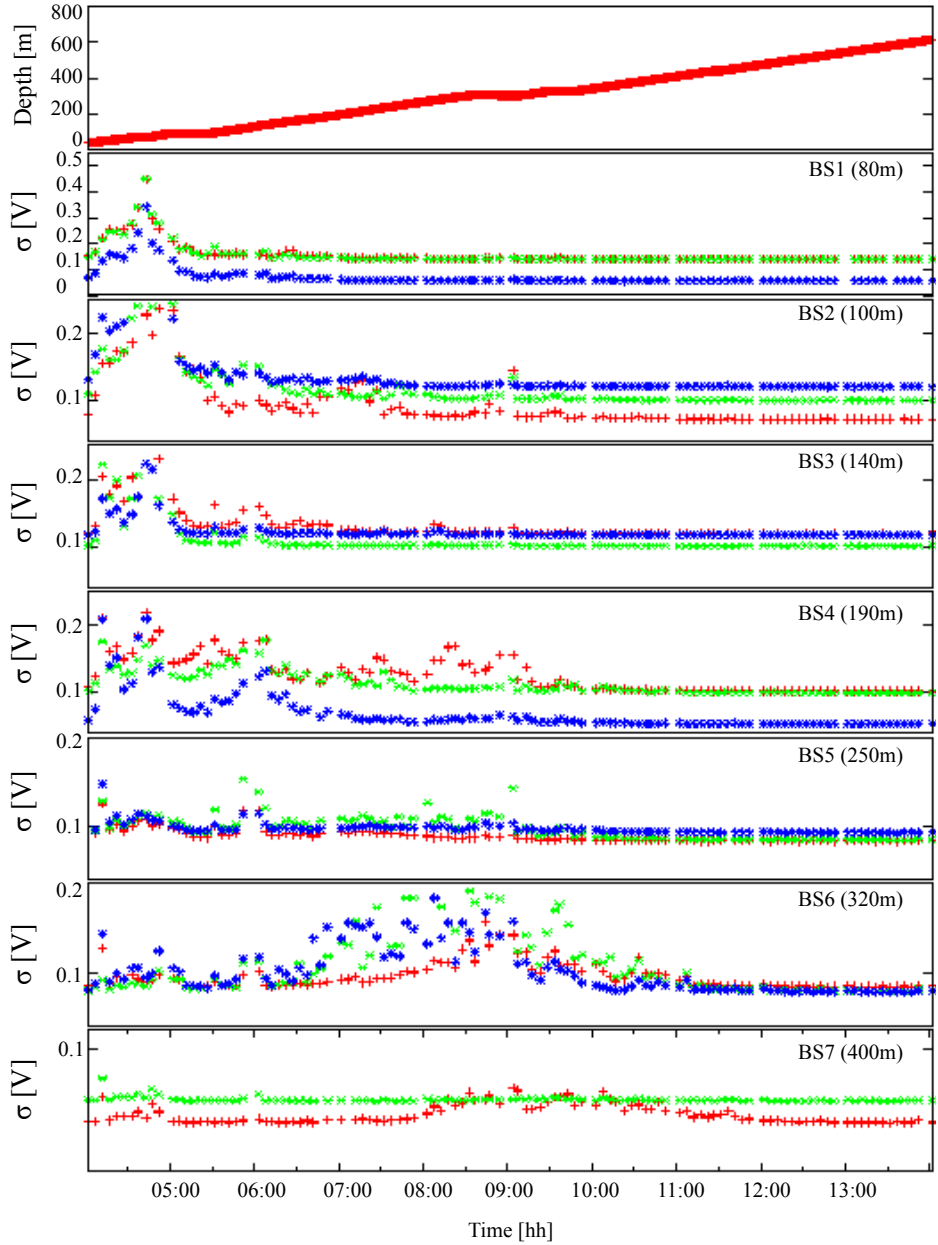}
\end{center}
\caption[Evolution of the SPATS noise levels for String\,B during
  drilling at IceCube hole 64.]{Evolution of the SPATS noise levels
  for String\,B during drilling at IceCube hole 64, at a distance of
  roughly 128\,m. The upper plot shows the depth of the IceCube
  hot-water drill versus time. The width of the Gaussian fitted to the
  histogram of the ADC voltages is shown for all available sensor
  channels.}
\label{fig:drillnoise}
\end{figure}

Since the observed noise in the SPATS sensors is Gaussian, the noise
levels in the ice can be monitored by looking at the evolution of the
standard deviation of the Gaussian ADC-count distributions for each
channel. The SPATS noise levels are found to be very stable. One
exception to this is that noise levels were observed to increase
during the IceCube drilling season. In fact, it was shown that the
drill can clearly be heard by the SPATS sensors if it is close
enough. Fig.\,\ref{fig:drillnoise} shows the width of the ADC-count
distributions for all available sensor channels of String\,B for a
period during which the IceCube drill was passing by at the depths of
each of the stages.

Currently the SPATS data-taking is concentrating on transient
data-taking: 45 minutes of each hour are dedicated to transient
data-taking, the remaining 15\,min are reserved for noise data-taking
and system monitoring. The thresholds for each channel that takes part
in transient data-taking are chosen so that the total data-rate for
SPATS remains under 150\,MB/day under normal trigger circumstances.
In total, over 700\,GB of data has been accumulated over the past
$\sim$4 years.

\section{Calibration and tests}
Prior to deployment the sensitivity in water and equivalent self-noise level of all sensors of strings A, B, and C 
were measured. At the South Pole the sensors are exposed to low temperatures (-50 C), increased static pressure, 
and a different acoustic coupling from the medium to the sensor. The change of sensor sensitivity with static pressure and 
temperature has been studied in dedicated laboratory experiments. We are currently investigating how the sensitivity, as determined in water, 
can be transferred to the in-situ conditions in the Antarctic ice.

A present estimate for the average SPATS sensor sensitivity in ice uses information from laboratory studies of temperature 
and pressure dependence on this quantity (see section 7.4) as well as data from the noise behaviour during sensor 
freeze-in at the South Pole. The derived value has however a big uncertainty. With a few sensor channels calibrations have 
been performed in a laboratory ice tank~\cite{meures:2010}. The data show that the frequency dependence of the sensitivity changes due to 
damping of mechanical resonances. Results of forthcoming new measurements of this type will hopefully allow to reduce the error on 
the sensor sensitivity determination in ice considerably.

\subsection{SPATS sensor calibration in water}

The SPATS sensors are calibrated relative to a commercial
hydrophone\footnote{SensorTech SQ03} with known spectral
sensitivity. This reference hydrophone has a rather flat sensitivity
curve for frequencies from $10 \munit{kHz}$ to $80 \munit{kHz}$ with a
mean sensitivity of $(-167.5 \pm 0.3) \munit{dB} \munit{re.}  1
\munit{V} / \mathrm{\mu Pa}$.

Ring shaped piezo ceramic elements, similar to the SPATS transmitters,
were used to generate a broadband pulse in a water tank at the {\em
  Hamburgische Schiffbau-Versuchsanstalt} (HSVA, Hamburg Ship Model
Basin). The available water volume of $12 \times 10 \times 5
\munit{m}^3$ is sufficiently large to allow clear separation between
direct signals and wall reflections. The acoustic devices were mounted
at a depth of $2 \munit{m}$ and the spacing between the transmitter
and the sensor was $1.03 \munit{m}$, in order to avoid near field
effects. The water temperature was $0.5^\circ \munit{C}$ and the
salinity was $7 \munit{ppt}$. Acoustic pulses were recorded with the
reference hydrophone and the frequency spectrum of the pressure wave
was derived. The SPATS sensor modules were positioned in the same
acoustic field. The sensor channel under investigation always pointed towards
the transmitter.

\begin{figure}
  \centering \includegraphics[height=4.0cm]{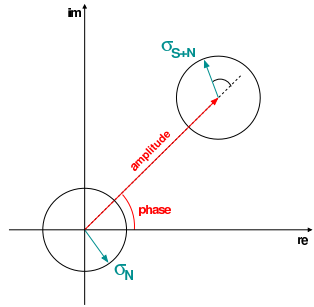}
  \caption{Schematic of noise contributions in the complex plane with
    signal amplitudes $A$ and phase $\phi$}
  \label{fig:noise_subtraction_schematic}
\end{figure}

\begin{figure}
  \centering
  \subfigure[]{\includegraphics[height=5.0cm]{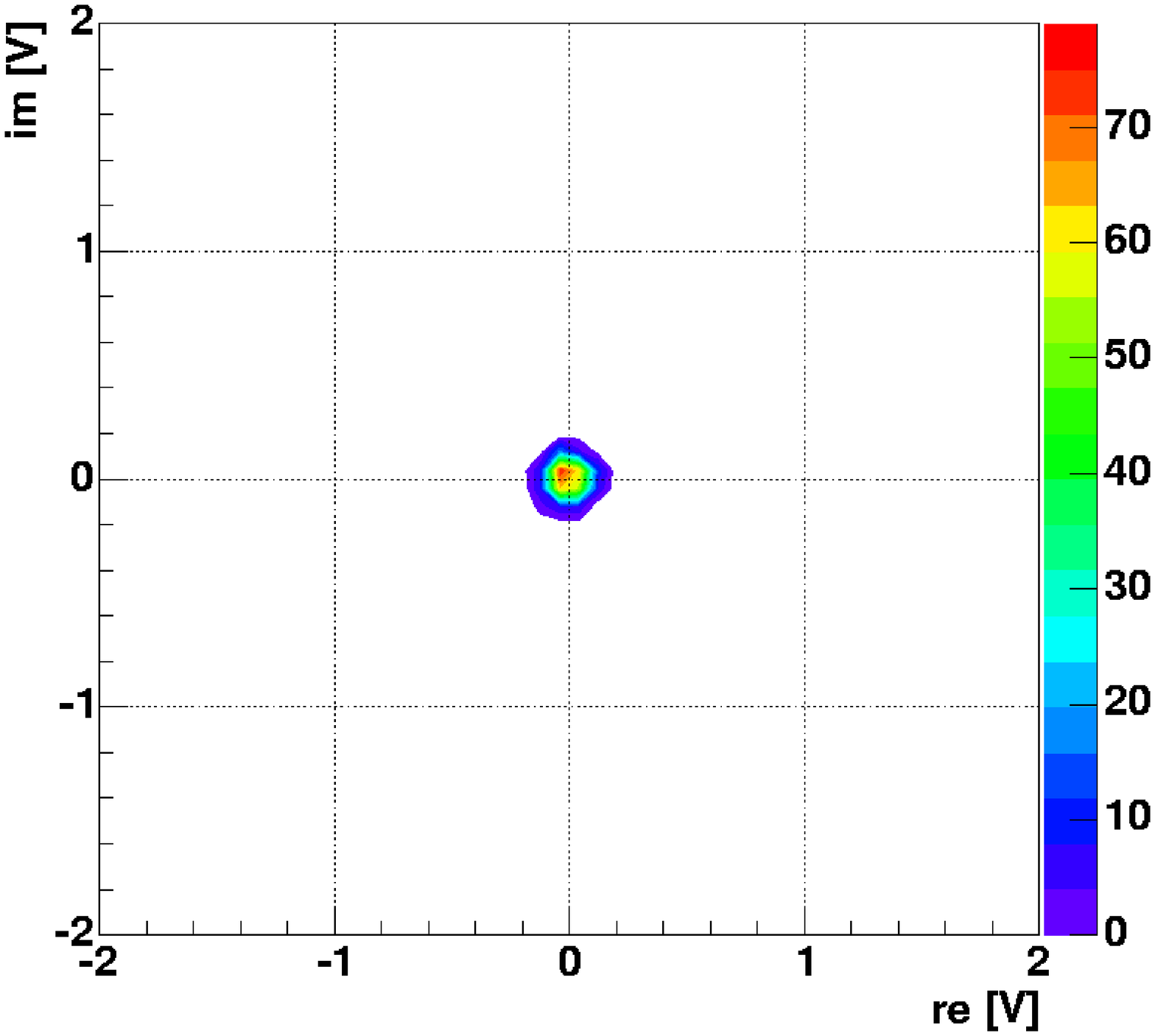}}
  \subfigure[]{\includegraphics[height=5.0cm]{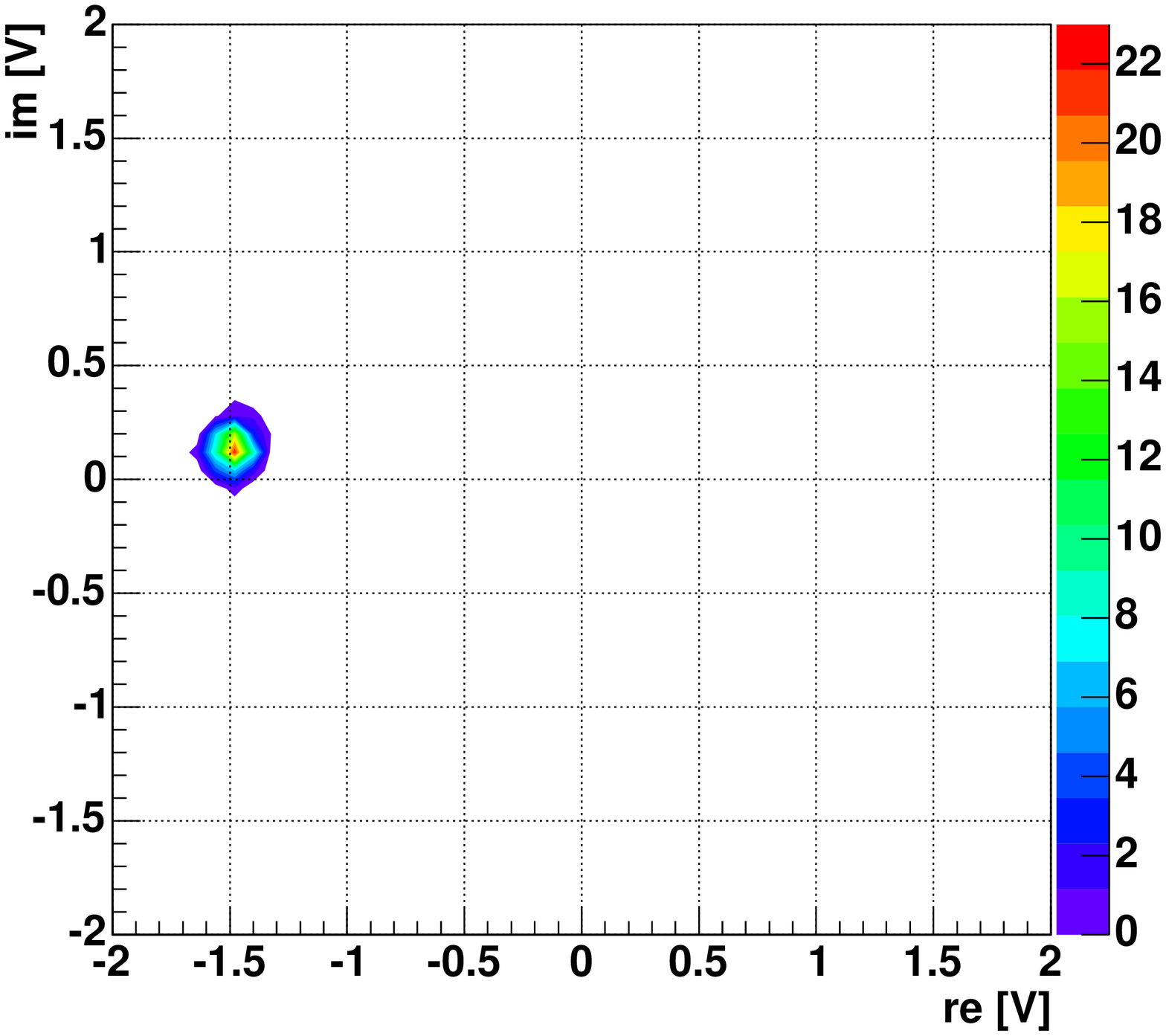}}
  \caption{Distribution of Fourier coefficients for a single sensor
    channel and one exemplary frequency. Off-signal (noise)
    contributions (a) and signal contributions (b)
    are shown. No significant contribution from pressure pulse
    variations is observed.}
  \label{fig:noise_subtraction}
\end{figure}

\begin{figure}
  \centering 
  \includegraphics[height=5.0cm]{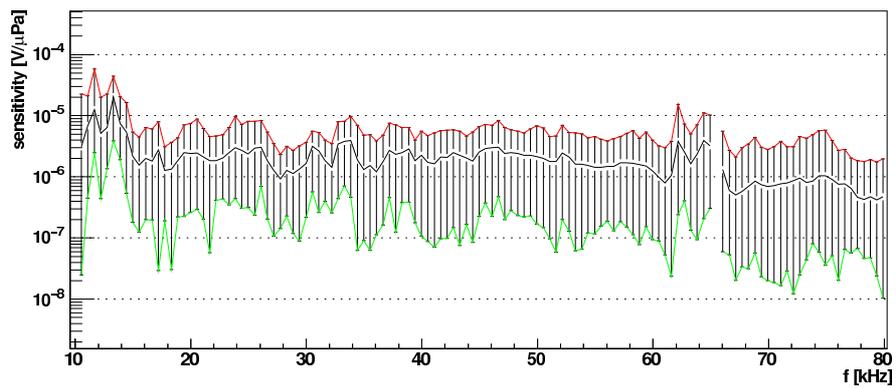}
  \caption{Compilation of all sensor channel sensitivities. The upper
    and lower lines indicate the highest and lowest sensitivities
    respectively; the black middle line corresponds to the mean value.}
  \label{fig:sensitivities}
\end{figure}

In order to cancel the expected contributions of background noise, $100$
pulses were recorded for all channels. The uncertainty on the
resulting signal frequency spectrum was calculated by dividing each
recorded signal into one signal region and four off-signal regions. A
discrete Fourier transform of the received signals results in fixed
amplitudes $A_i$ and phases $\phi_i$ for the Fourier coefficients,
which are Gaussian smeared by noise contributions
(see Fig.~\ref{fig:noise_subtraction_schematic}).  The width of the
two-dimensional Gaussian $\sigma_{\mathrm{signal}}$ consists of noise
contributions $\sigma_{\mathrm{noise}}$ and variations of the pressure
pulse shape $\sigma_{\mathrm{pulse}}$. From the difference of
$\sigma_{\mathrm{noise}}$ and $\sigma_{\mathrm{signal}}$ it was clear
that there was no significant contribution from pressure pulse
variation (see~Fig.~\ref{fig:noise_subtraction}). The signal
amplitudes were derived from the Fourier coefficient distributions and
the sensitivity spectra for all channels were obtained as the ratio of
the amplitude spectra and the previously determined pressure wave
frequency spectrum. The range of measured sensitivities for all SPATS
sensors is shown in Fig.~\ref{fig:sensitivities}. The mean sensitivity
averaged over all sensors and frequencies is $2.6 \cdot 10^{-6}
\munit{V} / \mathrm{\mu Pa}$ equal to $-112 \munit{dB} \munit{re.}  1
\munit{V} / \mathrm{\mu Pa}$.

\subsection{Sensor equivalent self-noise}

\begin{figure}
  \centering \includegraphics[height=5.0cm]{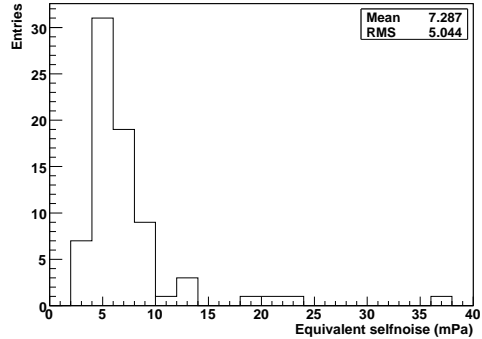}
  \caption{Compilation of equivalent self-noise levels integrated from
    $10 \munit{kHz}$ to $50 \munit{kHz}$ for all SPATS channels.}
  \label{fig:selfnoise}
\end{figure}

For each SPATS sensor channel self-noise measurements were
performed. External acoustic and electromagnetic interferences was
minimized by appropriate shielding. Using the sensor sensitivity, the
self-noise frequency spectrum was transformed to an equivalent
self-noise spectrum. Integration of this equivalent self-noise
spectrum over the frequency range of interest resulted in the
equivalent self-noise level. This was used as a measure for the
threshold above which acoustic waves were detectable by the sensor. In
Fig.~\ref{fig:selfnoise} the equivalent self-noise level integrated
from $10 \munit{kHz}$ to $50 \munit{kHz}$, the frequency range of
interest for acoustic neutrino detection, for all SPATS channels is
shown.

\subsection{Transmitter directivity calibration}
\label{sec:transdircal}
\begin{figure}[ht]
  \centering
  \subfigure[]{\label{fig:trans-azi}\includegraphics[height=5.0cm]{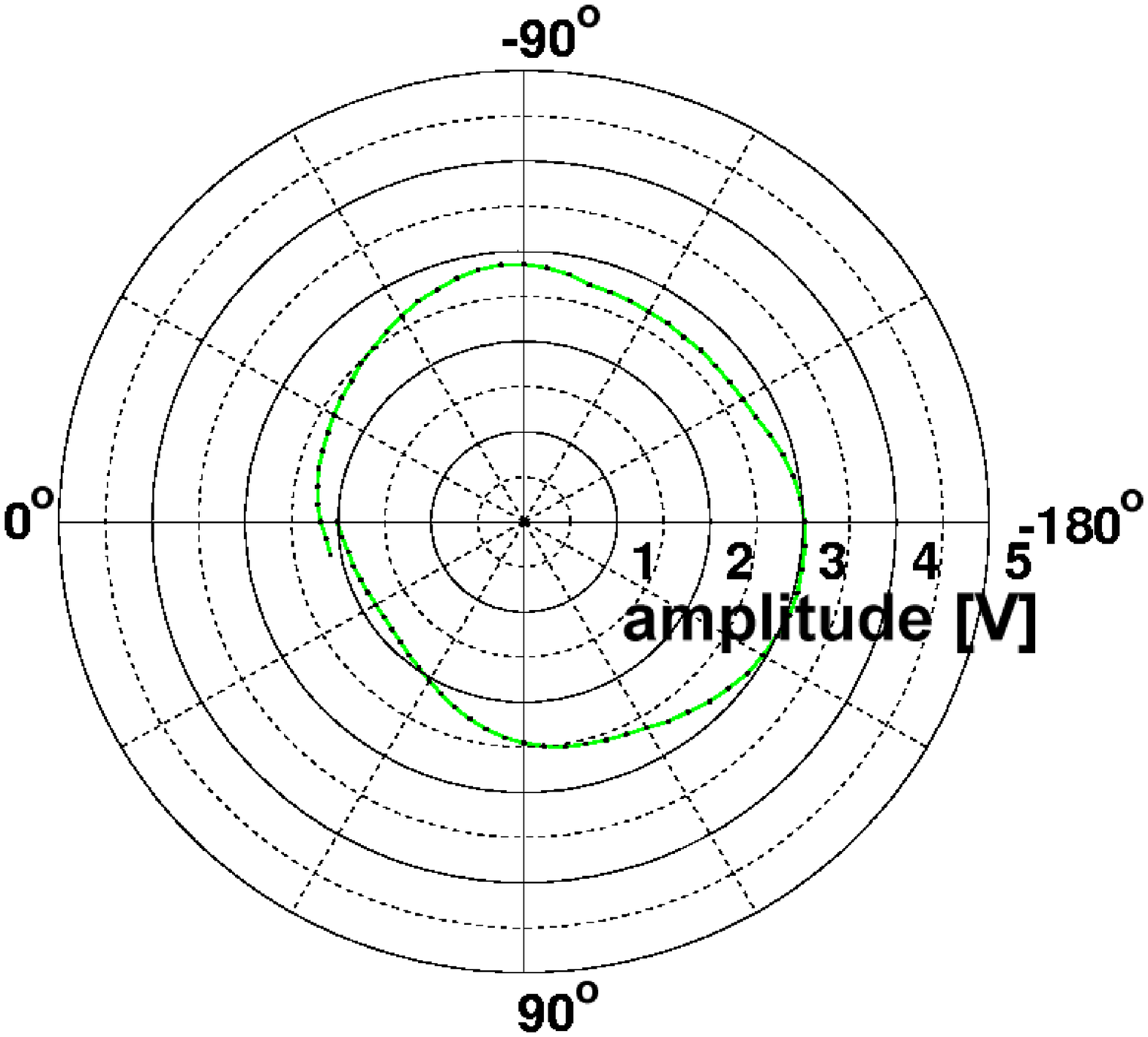}}
  \hspace{0.5cm}\subfigure[]{\label{fig:trans-pol}\includegraphics[height=5.0cm]{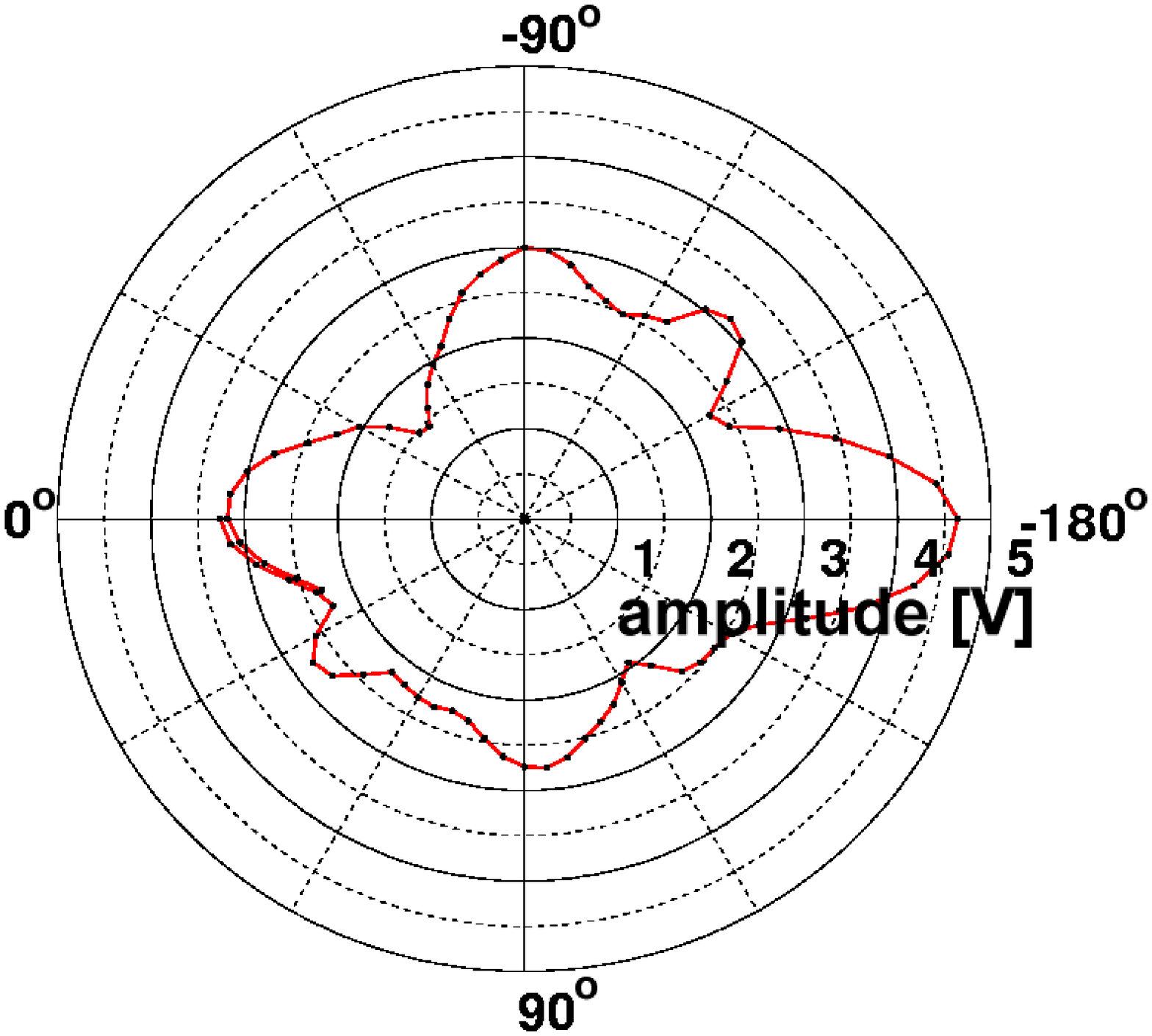}}
  \vspace{0.3cm}
  \subfigure[]{\label{fig:Setup_ringtransmitter_azi}\includegraphics[height=4.5cm]{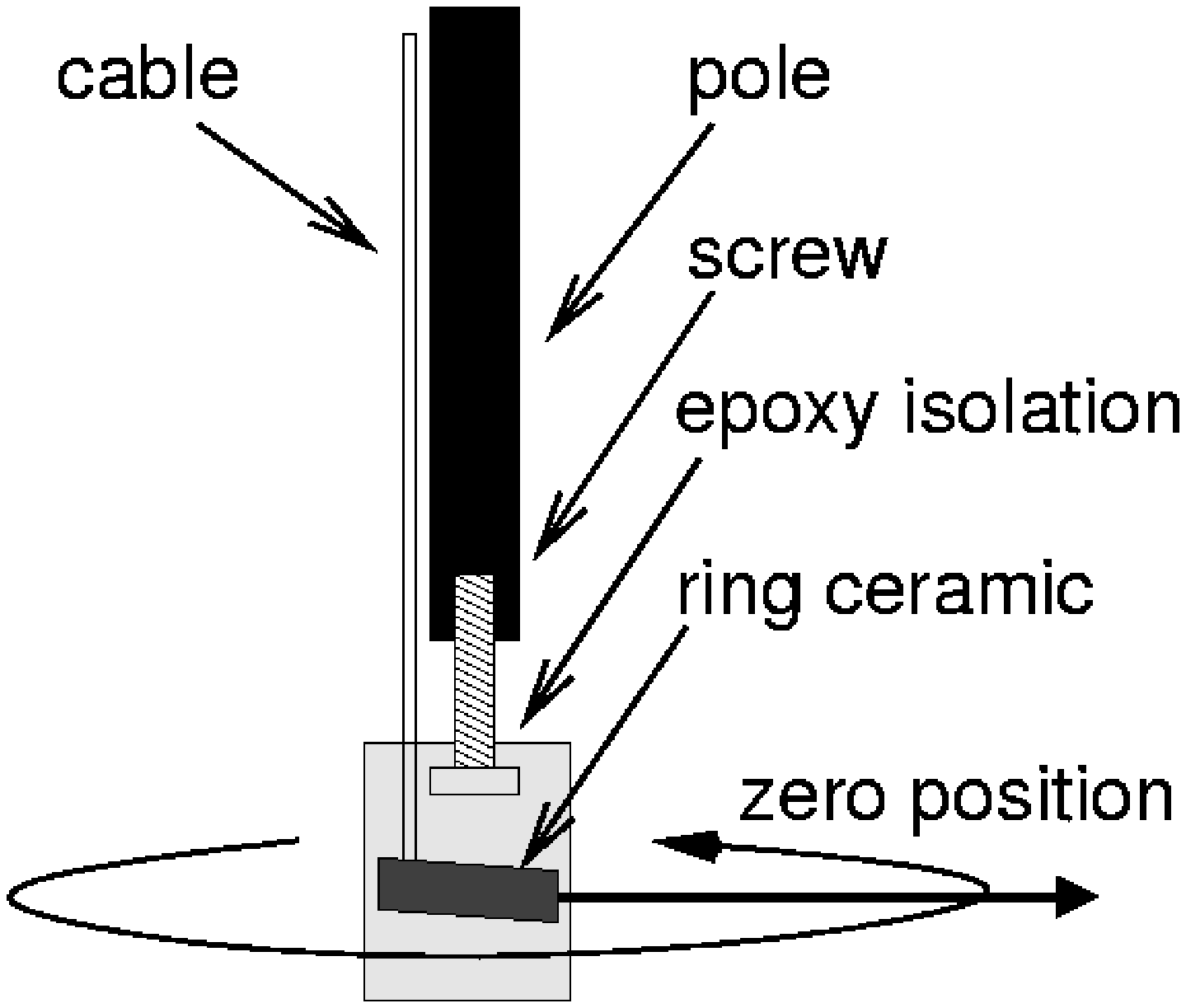}}
  \caption{\label{fig:trans} Received pulse amplitudes as function of
    transmitter \subref{fig:trans-azi} azimuthal and
    \subref{fig:trans-pol} polar orientation. \subref{fig:Setup_ringtransmitter_azi} shows the setup for the azimuthal measurement.}
\end{figure}

Azimuthal isotropic emission is the motivation for the use of ring
shaped piezo ceramics. The actual emission directivity of a ring
shaped piezo ceramic inside epoxy was measured in azimuthal and polar
direction. For measuring polar variations, the piezo ceramic was
turned around an axis perpendicular to the ring symmetry axis. The
pulses were recorded by the reference hydrophone at a distance of $1
\munit{m}$. Figure \ref{fig:trans} shows the received pulse amplitudes
as function of azimuth and polar angle. Figure \ref{fig:trans}\subref{fig:Setup_ringtransmitter_azi} shows a 
schematic for the azimuthal directivity measurement. It illustrates how the positioning of the piezo inside 
the epoxy could tilt the piezo away form the azimuthal plane. 

During deployment in South Pole ice, there is no control over the
azimuthal orientation. Amplitude variations of around $40$\% are
expected due to this systematic effect. 

The variation of the amplitude outside the horizontal plane in the ice is defined by the polar emission pattern.
Despite exhibiting a large variation over the total range, on small scales
the polar emissivity is smooth. For repetitive measurements that sample the polar range with a variation of no more than $\pm 10^\circ$, 
the amplitude variation will be below $10$\%.

\subsection{Pressure and temperature dependence}
\label{sec:pressuretest}
\subsubsection{Pressure}
The deepest SPATS stage is deployed at a depth of 500\,m and the
maximum ambient pressure is expected to be 100\,bars. This is due to
the fact that the water inside the IceCube hole refreezes first at the
surface, creating over-pressure during the freeze-in. The SPATS steel
pressure housings have been tested at static pressures up to 120\,bar, which
is roughly the equivalent of 1200\,m of water depth. The housings are not
expected to be deformed by more than roughly 50\,$\mu$m at a tension
of 180\,N/m$^{2}$. It is possible that a slight deformation gives rise
to a change in sensitivity. For example, the preload on a
piezo-ceramic element could change due to deformation of the steel
housing.

\begin{figure}
 \centering
 \subfigure[]{\label{fig:PressDep}\includegraphics[height=4.7cm]{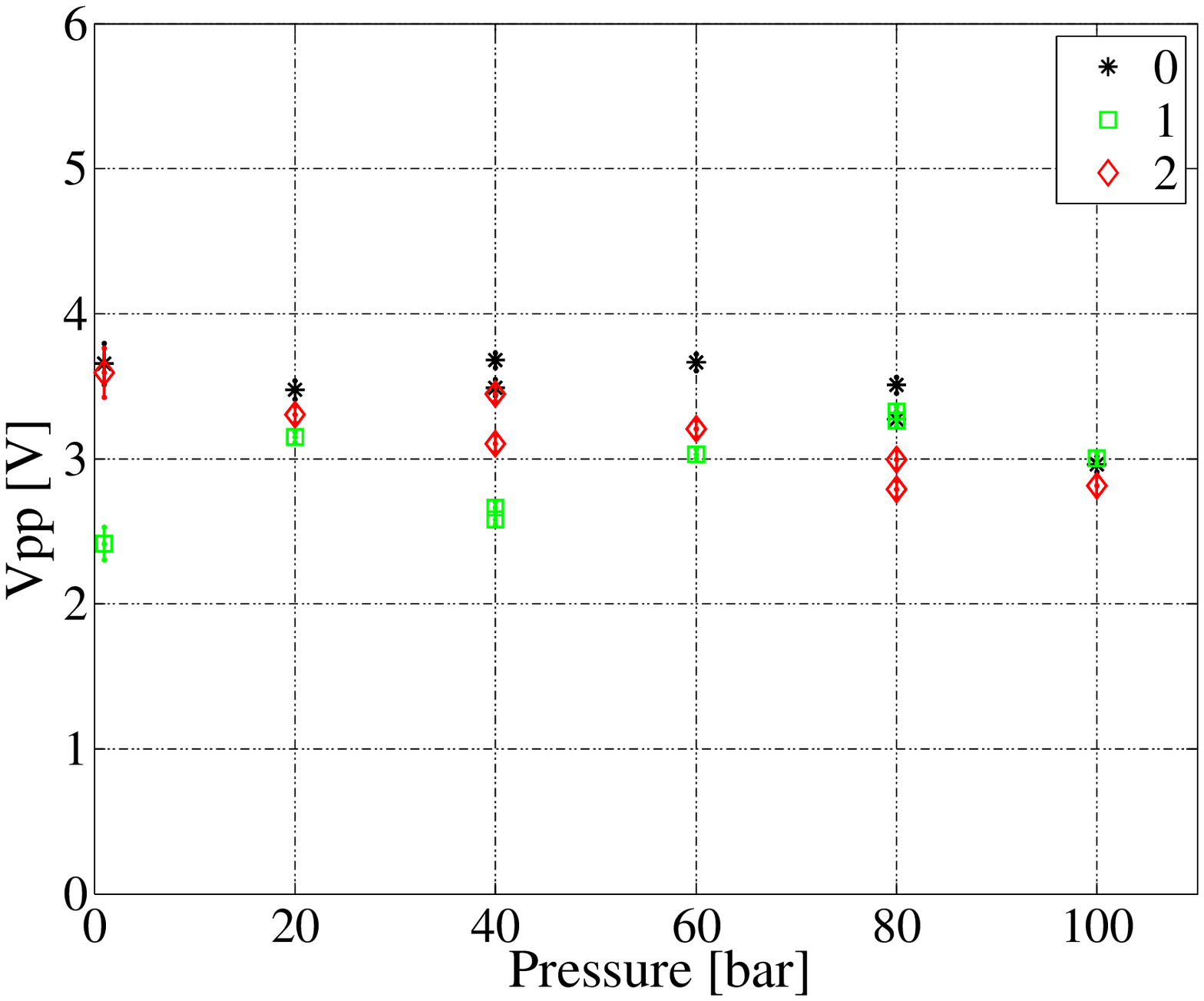}}
 \subfigure[]{\label{fig:TempDep}\includegraphics[height=4.7cm]{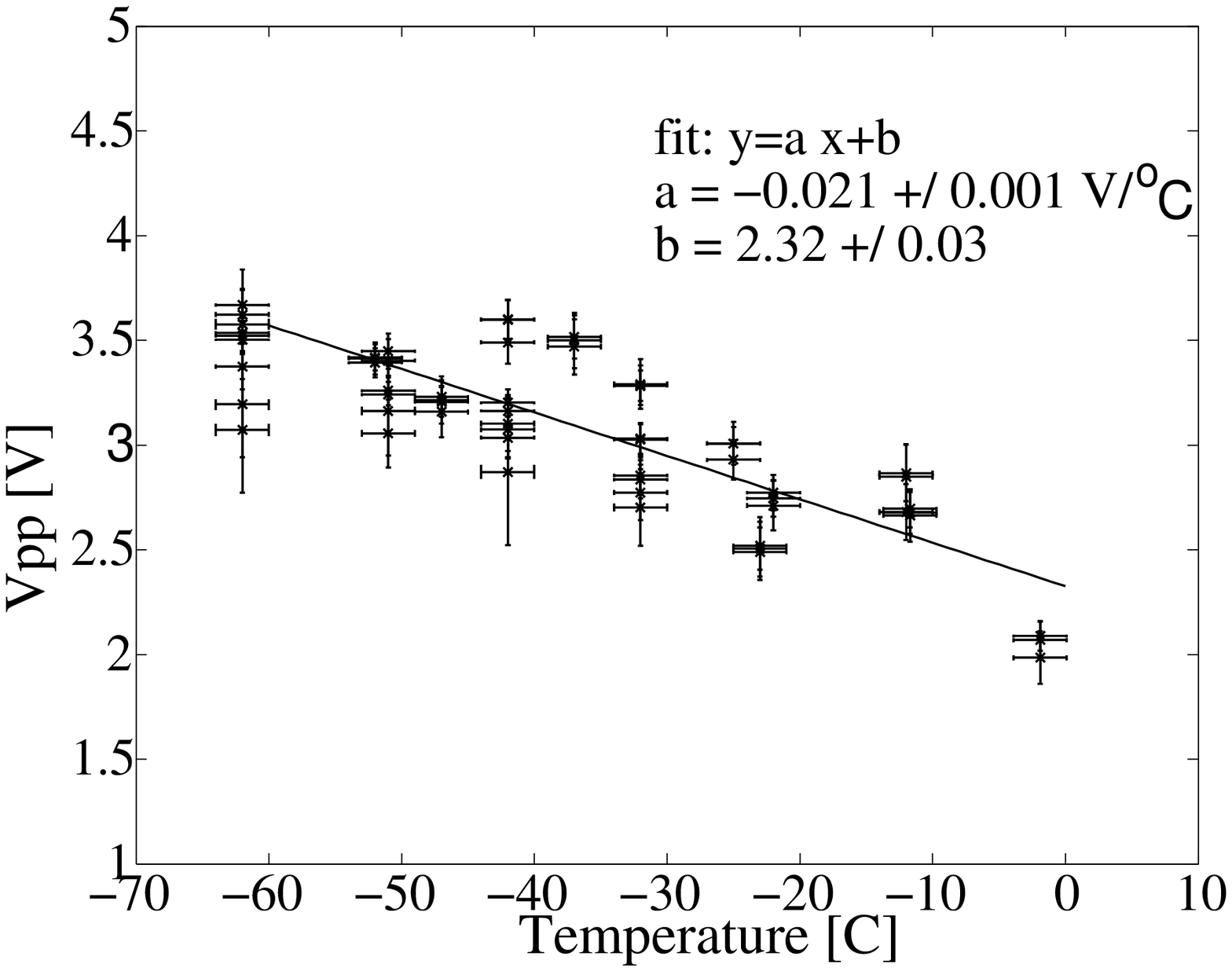}}
 \caption[]{\label{figure:sensdep}(a): The peak-to-peak voltage (Vpp)
    as a function of applied static pressure for the three channels of
    the same SPATS sensor listening to a transmitter external to the
    pressure vessel. (b): Vpp as a function of temperature for the
    three channels of the same SPATS sensor in air with a linear fit
    to the data (black line).}
\end{figure}

In order to quantify the change in sensitivity due to the increase of
ambient pressure, two pressure test sequences were done at Uppsala
University, in Sweden, in August~2008 and February~2009. A large
(40.5\,cm inner diameter) pressure vessel was used in which SPATS or
HADES sensors were consecutively installed. A commercial transmitter
(SQ9) was mounted outside the pressure vessel for the February~2009
tests. This data-set consisted of pressure dependent
measurements of the transmitter pulse at different operating voltages.
The transmitter was pulsed with a single-cycle 20\,kHz sine wave. These signals were
recorded by the sensors inside the pressure vessel at a maximum
pressure of 100\,bar. The typical step in pressure was 20\,bar. The
recorded peak-to-peak voltage (Vpp) served as a measure of sensitivity
at the various pressure levels. Figure~\ref{fig:PressDep} shows the
measured Vpp for the three channels of a SPATS sensor as a function of
ambient pressure: no systematic variation with pressure, common to all
three channels, is visible. The complete pressure data set indicated
that the variation of sensitivity of a SPATS sensor channel with
static pressure was less than 30\% between 1\,bar and 100\,bar.

\subsubsection{Temperature}
The mean ambient temperature in the South Pole ice in the upper few
hundred meters is expected to be around --50\,$^{\circ}$C. Several
temperature tests have investigated the temperature dependence of the
SPATS sensors. To this end, industrial freezers capable of reaching
temperatures below --60\,$^{\circ}$C were used.  In an experimental
verification, performed in 2009, an ITC1001 transmitter was pulsed in
air while being suspended near the freezer at the DESY laboratory in
Zeuthen, Germany. A SPATS sensor was positioned on a support at the
bottom of the freezer. The position of the sensor was not changed
between the different measurements. The temperature for each
measurement was taken with a thermocouple and a digital thermometer
which were both in contact with the steel housing of the sensor. The
difference between the two temperature measurements was typically
2\,$^{\circ}$C which was used as the error. The peak-to-peak
amplitudes were extracted as the mean of 100 events with the standard
deviation as the statistical error. Figure~\ref{fig:TempDep} shows the
extracted Vpp as a function of temperature for the three channels of a
same SPATS sensor module. A linear fit was made to the data and it was
concluded that there was a gain in the sensitivity of $1.5\pm 0.2$
when the temperature was lowered from 0\,$^{\circ}$C to
--50\,$^{\circ}$C.

\subsection{Abisko lake test}
A SPATS full system test was performed in lake Tornetr\"{a}sk in the
Abisko national park in northern Sweden in April 2006. At the test
location the lake was between $40 \munit{m}$ and $60 \munit{m}$ deep
and was covered with about $90 \munit{cm}$ of ice. Sensors and
transmitters were deployed and read out using a portable DAQ system. Holes
were drilled in the ice on a straight line from north to south, where the maximum
distance was about $800 \munit{m}$, and two holes were located at $400
\munit{m}$ west and $400 \munit{m}$ east. The goals of this test were
threefold: firstly the transmitters were tested to investigate variations
between different modules and to demonstrate that the range is
sufficient to meet SPATS requirements. Secondly, the performances of
the sensor modules were compared to that of the reference hydrophone
and the direction of the pulse was reconstructed using two channels
per module.  Finally, the DAQ software and hardware were tested under
real deployment conditions.  For each measurement of a transmitter
pulse at a certain distance and depth, $10$ events were recorded by
the reference hydrophone or one of the SPATS sensors. Different
effects contributed to shifts in the arrival time of the
signal. Underwater currents make the devices swing or torsion in the
support rope makes them spin. For analysis, the signals were therefore
shifted in time so that the first amplitude maxima coincide, before
extracting the mean amplitudes.  Lake Tornetr\"{a}sk was found to be a
silent testing environment under good weather conditions with a stable and low background noise
level of about $120 \munit{mV}$ for the SPATS sensors, excluding
occasional snow scooters and strong winds.

\begin{figure}
  \centering
  \subfigure[]{\label{fig:abis-tamp}\includegraphics[height=4.3cm]{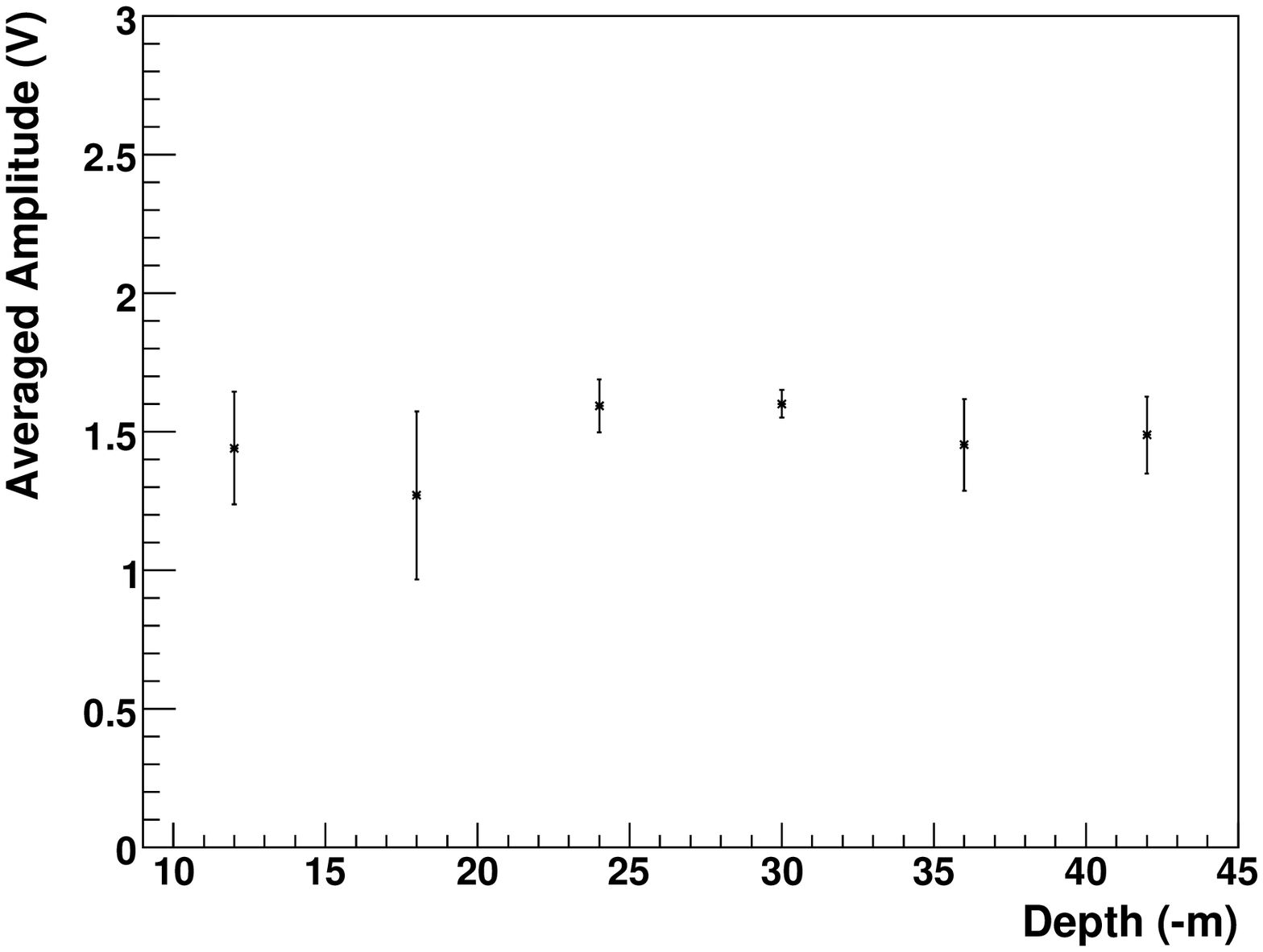}}
  \subfigure[]{\label{fig:abis-800}\includegraphics[height=4.3cm]{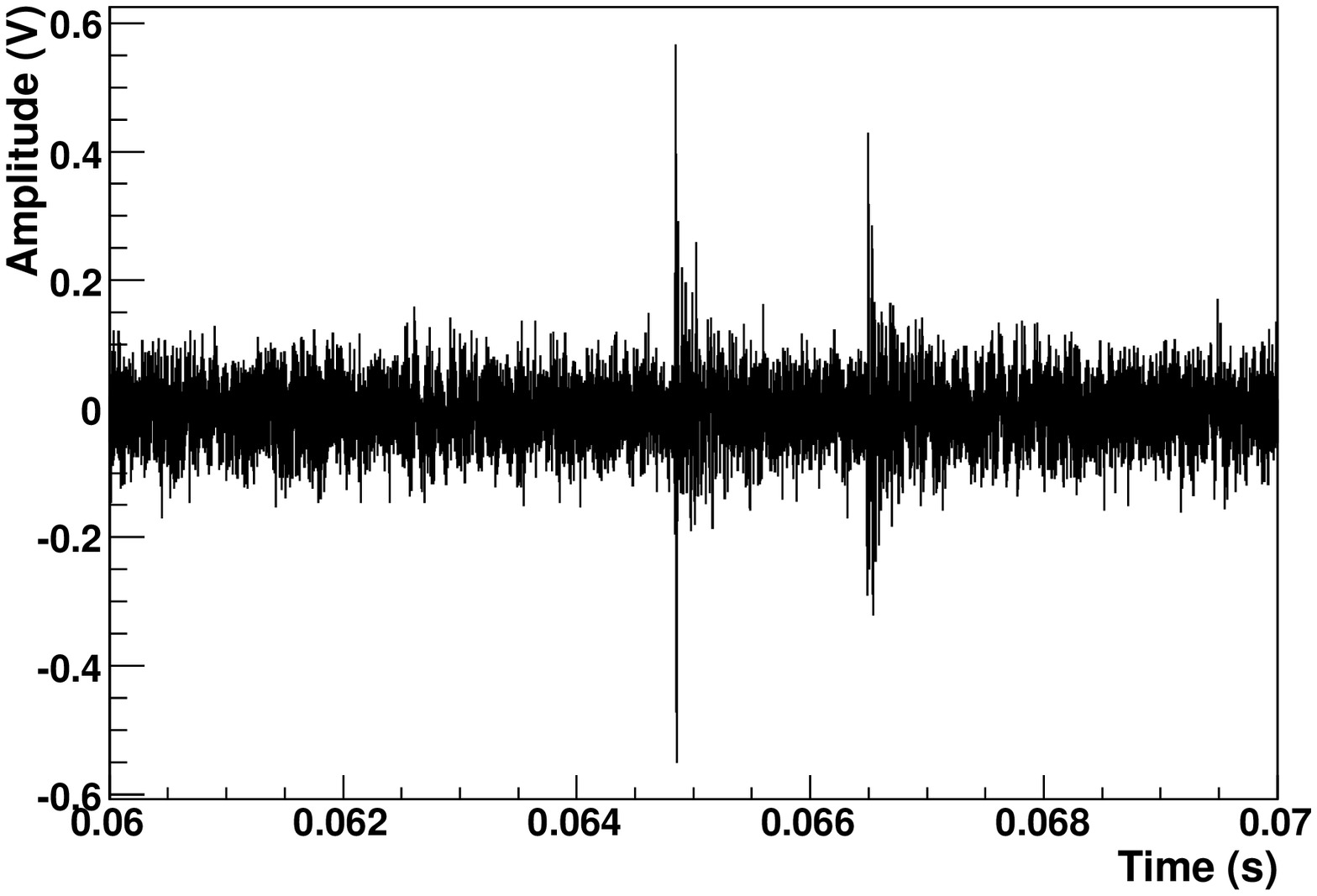}}
  \caption{\label{figure:abis1}Results from the Abisko transmitter
    test: \subref{fig:abis-tamp} variation in amplitude for one
    transmitter at different depths, and \subref{fig:abis-800}
    transmitter signal as seen by a SPATS sensor at 800\,m distance.}
\end{figure}

At first, a SPATS sensor was placed at a depth of $30 \munit{m}$ and
different combinations of transmitter modules and ring shaped piezo
ceramics were lowered to varying depths in a hole at a distance of
$400 \munit{m}$. The amplitudes of the registered pulses were
compared; an example can be found in Fig.~\ref{fig:abis-tamp}. Signals
from all tested transmitters are clearly visible at a distance of $400
\munit{m}$. The observed variations in amplitude lie within the
expectations and are due to the uncertainties on azimuthal and polar
orientation of the transmitter.  A maximum distance of $800 \munit{m}$
between transmitter and sensor was then achieved, where the available
cable length was the limiting factor for the range test. In
Fig~\ref{fig:abis-800}, the recorded signal with a signal to noise ratio
of $5$, is shown. The transmitter was positioned at a depth of $20
\munit{m}$. There is a second pulse visible approximately $1
\munit{ms}$ after the signal. A calculation taking into account the
speed of sound profile showed that the second pulse is a reflection
originating from the ice surface. Assuming an acoustic amplitude
attenuation length of $1 \munit{km}$, an extrapolation from the signal
to noise ratio gives an expected signal to noise ratio of 1 for the SPATS transmitter in
water of $1800 \munit{m}$.

\begin{figure}
   \centering
   \subfigure[]{\label{fig:abis-hyd}\includegraphics[height=4cm]{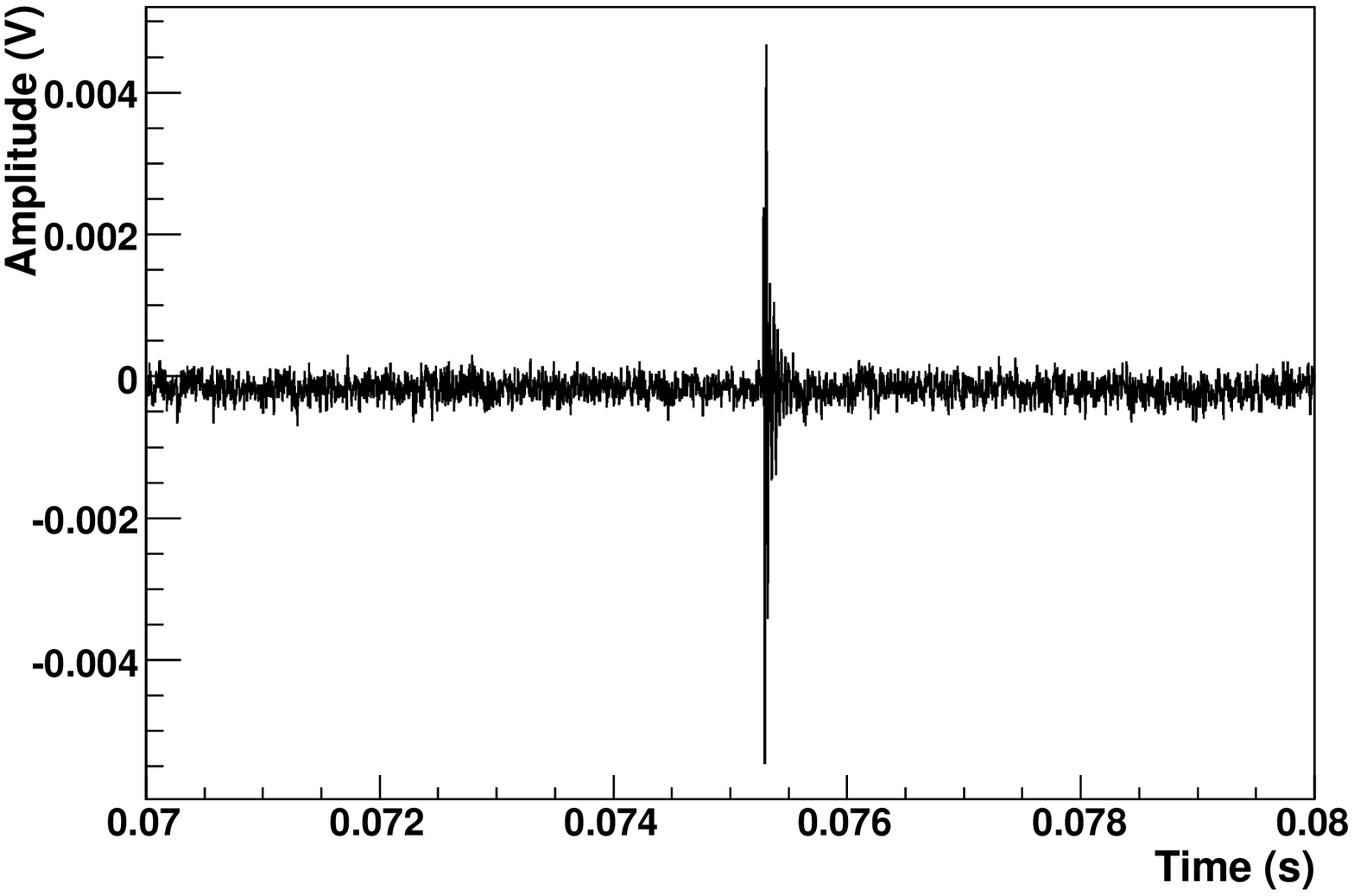}}
   \subfigure[]{\label{fig:abis-100}\includegraphics[height=4cm]{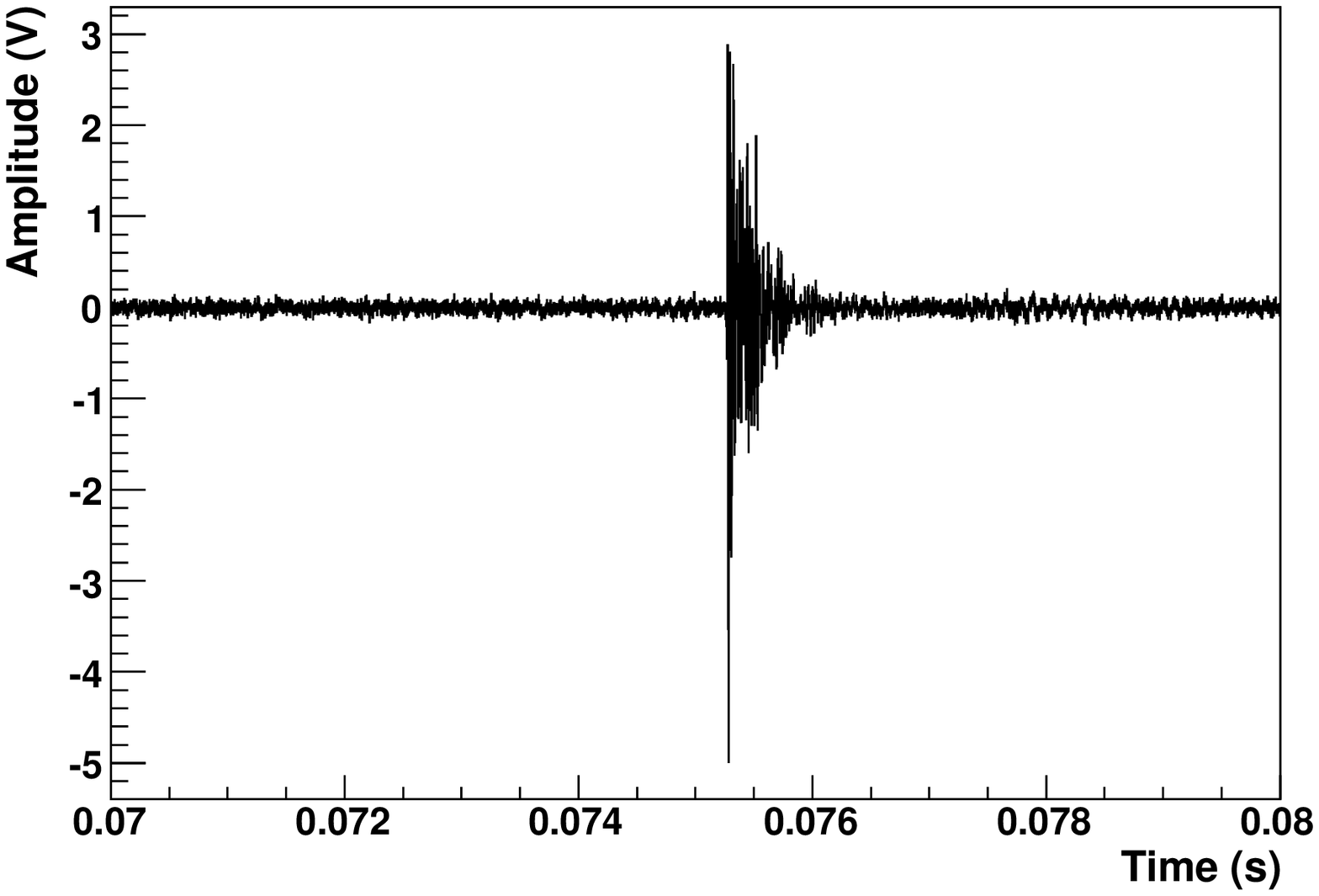}}
   \caption{\label{figure:abis2}Sensor results from the Abisko lake
     test: comparison between the signals received by (a) the commercial
     hydrophone and (b) SPATS sensor at a distance of 100\,m to the transmitter.}
 \end{figure}

 Second, the difference in performance of the SPATS sensor and the
 reference hydrophone becomes clear in Fig.~\ref{figure:abis2}, where
 the significant difference in scale should be noted. The transmitter
 was placed at a distance of $100 \munit{m}$ and a depth of $30
 \munit{m}$ after which the hydrophone and SPATS sensor were
 successively lowered to a depth of $16 \munit{m}$. The signal as
 recorded by the sensor is much stronger than that of the commercial
 hydrophone. In fact, the hydrophone was incapable of detecting a
 transmitter signal at a distance $400 \munit{m}$ at the maximum gain
 setting.

Finally, the SPATS sensor was placed at a depth of $30 \munit{m}$ and
a transmitter was lowered to the same depth at positions $400
\munit{m}$ north, west, south and east of the sensor location. The
three sensor channels are each separated by $10.5 \munit{cm}$ of
steel. The orientation of the sensor module and thus the
directional information was obtained by reading out two channels for
each position of the transmitter. The difference in arrival times between different channels could
be observed at the maximum sampling frequency of $1.25
\munit{MHz}$. This way, the sensor channel that was closest to the source could be identified.

\section{Conclusion and outlook}
The SPATS detector was successfully deployed in the Antarctic ice at
the Geographic South Pole and has been recording data for over 5
years. All components of the detector were designed to perform under
the extreme temperatures and pressures. No major issues occurred after
installation and several unexpected power-outages have not affected
system performance.  Most of the science goals have been reached and
the physics results have both confirmed and challenged theoretical
predictions by producing the first ever experimental results for the
acoustic attenuation length~\cite{Abbasi:2010vt}, both the pressure and shear wave sound
speeds~\cite{Collaboration:2009sia} and the noise levels and transient sounds~\cite{noise} in the South Pole bulk ice. SPATS
is currently undergoing a system upgrade and continues to take
acoustic measurements.
 
\section{Acknowledgments}
We acknowledge the support from the following agencies: U.S. National
Science Foundation-Office of Polar Program, Swedish Research Council,
Swedish Polar Research Secretariat, and Knut and Alice Wallenberg
Foundation, Sweden; German Ministry for Education and Research (BMBF),
Deutsche Forschungsgemeinschaft (DFG), Germany; Fund for Scientific
Research (FNRS-FWO), Flanders Institute to encourage scientific and
technological research in industry (IWT), Belgian Federal Science
Policy Office (Belspo); M.~Ribordy acknowledges the support of the SNF
(Switzerland).

\bibliographystyle{elsart-num}
\bibliography{bib}

\end{document}